\documentclass[10pt,journal]{IEEEtran}
\IEEEoverridecommandlockouts
\usepackage{xcolor}
\usepackage{adjustbox}
\usepackage{subfigure}
\usepackage{soul}
\usepackage{mathtools}
\usepackage{balance} % For balanced columns on the last page
\usepackage{url}
\usepackage{slashbox}
\usepackage{graphicx}
\usepackage{caption}
\usepackage{multirow,booktabs} %for the table
\usepackage{array}
\usepackage{makecell}
\usepackage{tikz}

\usepackage{colortbl}

\newcommand{\ie}{{\em i.e., }}
\newcommand{\eg}{{\em e.g., }}

\newcommand{\myverb}{\fontsize{10}{48}\usefont{OT1}{lmtt}{b}{n}\noindent }

\newcolumntype{L}[1]{>{\raggedright\let\newline\\\arraybackslash\hspace{0pt}}m{#1}}
\newcolumntype{C}[1]{>{\centering\let\newline\\\arraybackslash\hspace{0pt}}m{#1}}
\newcolumntype{R}[1]{>{\raggedleft\let\newline\\\arraybackslash\hspace{0pt}}m{#1}}

\ifCLASSINFOpdf
\else
\fi

\begin{document}

%make title bold and 14 pt font (Latex default is non-bold, 16 pt)

%\title{Enterprise DNS Forensics Analysis: \\Host Mapping and Traffic Health Tracking}

\title{Analyzing Enterprise DNS Traffic \\to Classify Assets and Track Cyber-Health}

\author{
	Minzhao~Lyu,
	Hassan~Habibi~Gharakheili,
	Craig~Russell,		
	and~Vijay~Sivaraman%,~\IEEEmembership{Member,~IEEE}% <-this % stops a space
	
	\IEEEcompsocitemizethanks{
		\IEEEcompsocthanksitem	M. Lyu is with the School of Electrical Engineering and Telecommunications, University of New South Wales, Sydney, NSW 2052, Australia, and CSIRO's Data61, Sydney, NSW 2015, Australia (e-mail: minzhao.lyu@unsw.edu.au).
%		\protect\\
		\IEEEcompsocthanksitem H. Habibi Gharakheili, C. Russell and V. Sivaraman are with the School of Electrical Engineering and Telecommunications, University of New South Wales, Sydney, NSW 2052, Australia (e-mails: h.habibi@unsw.edu.au, craig.russell@unsw.edu.au, vijay@unsw.edu.au). 
%		\protect\\
		
		\IEEEcompsocthanksitem This article is an extended and improved version of our paper presented
		at the Passive and Active Measurement (PAM) 2019 conference \cite{MLyuPAM2019}.	
}
}

%\markboth{IEEE Transactions on Information Forensics and Security}%
%{M. Lyu \MakeLowercase{\textit{et al.}}: Analyzing Enterprise DNS Traffic to Classify Assets and Track Cyber-Health}

\IEEEtitleabstractindextext{%
	\begin{abstract}
		%word limit is 250 words
		
		The Domain Name System (DNS) is a critical service that enables domain names to be converted to IP addresses (or vice versa); consequently, it is generally permitted through enterprise security systems (\eg firewalls) with little restriction. This has exposed organizational networks to DDoS, exfiltration, and reflection attacks, inflicting significant financial and reputational damage. Large organizations with loosely federated IT departments (\eg Universities and Research Institutes) often do not even fully aware of all their DNS assets and vulnerabilities, let alone the attack surface they expose to the outside world.
		
		In this paper, we address the ``DNS blind spot'' by developing methods to passively analyze live DNS traffic, identify organizational DNS assets, and monitor their health on a continuous basis. Our contributions are threefold. First, we perform a comprehensive analysis of all DNS traffic in two large organizations (a University Campus and a Government Research Institute) for over a month, and identify key behavioral profiles for various asset types such as recursive resolvers, authoritative name servers, and mixed DNS servers. Second, we develop an unsupervised clustering method that classifies enterprise DNS assets using the behavioral attributes identified, and demonstrate that our method successfully classifies over 100 DNS assets across the two organizations. Third, our method continuously tracks various health metrics across the organizational DNS assets and identifies several instances of improper configuration, data exfiltration, DDoS, and reflection attacks. We believe the passive analysis methods in this paper can help enterprises monitor organizational DNS health in an automated and risk-free manner.
	\end{abstract}
%	\vspace{-3mm}
}

\maketitle
\IEEEdisplaynontitleabstractindextext
\IEEEpeerreviewmaketitle

\vspace{8mm}
\IEEEraisesectionheading{\section{Introduction}\label{sec:intro}}
\IEEEPARstart{E}{nterprise} networks are large in size with many thousands of connected devices and dynamic in nature as hosts come and go, and servers get commissioned and decommissioned to adapt to the organization's changing needs. Enterprise IT departments track such assets manually today, with records maintained in spreadsheets and configuration files (DHCP, DNS, Firewalls, etc.). This is not only cumbersome, but also error-prone and almost impossible to keep up-to-date. It is, therefore, not surprising that many enterprise network administrators are not fully aware of their internal assets \cite{SMarshall2013}, and consequently do not know the attack surface they expose to the outside world. The problem is even more acute in University and Research Institute campus networks for several reasons \cite{Deloitte2018}: (a) they host a wide variety of sensitive and lucrative data, including intellectual property, cutting-edge research datasets, social security numbers, and financial information; (b) their open-access culture, decentralized departmental-level control, as well as federated access to data makes them particularly vulnerable targets for unauthorized access, unsafe Internet usage, and malware; and (c) they typically have a high-speed network infrastructure that makes them an attractive launchpad for volumetric attacks on other entities.

DNS is a protocol of choice exploited by cyber-crimes and botnets as it can readily bypass firewalls and security middleboxes. Due to the open nature of DNS, it is common for organizations to apply few (if any) restrictions (\eg firewall rules) to DNS traffic. 
Thus, it is unsurprising to see the increasing frequency and quantity of malware compromised devices and attacks that leverage DNS protocol \cite{EfficientIP2018,JVijayan2018,CFachkhaNTMS2014,US-CERT2018,MAnagnostopoulosCS2013}, such as DDoS attack, DNS tunneling and sensitive data exfiltration.

Enterprises typically host various kinds of DNS assets. They will typically host a small number of recursive resolvers that proxy DNS requests from internal hosts to external DNS servers, and also cache results to reduce the number of external queries. Individual hosts may choose to over-ride the enterprise recursive resolvers, such as by manually changing their preferred resolver to a public one (such as Google's 8.8.8.8 and CloudFlare's 1.1.1.1), but in general, a majority of hosts will use the default recursive resolver provided by their organization. In addition, enterprises typically host a number of authoritative name servers to serve the various domains belonging to the organization. For example, organization-wide services (like email, VPN, etc.) may be managed by central IT. At the same time, each department may operate its own authoritative name server to resolve department-specific web pages. It is not uncommon for the various IT entities to operate in silos, often unaware of the assets being managed by the other. To make matters worse, on-campus retail stores (bookshops, food outlets, etc.) that lease connectivity from the campus may also be housing their own DNS assets, which are often poorly secured as they lack the skills.

Existing enterprise network security appliances such as border firewalls and intrusion detection systems do not provide fine-grained visibility of internal DNS-related hosts and assets, creating significant ``blind spots'' for the IT department \cite{CiscoASR2016}. While there is a significant body of existing academic research on DNS traffic analysis and DNS security, existing works either focus on forensic analysis of logs collected from DNS servers, such as recursive resolvers on the Internet \cite{XMATIFS2014,MAntonakakisSec2012,HGaoTNET2016,YChenDSN2014} and domain registrars \cite{SHaoCCS2016}, or concentrate on packet-level domain names \cite{YLeeCoNEXT2017,JAhmedIM2019,MAlmeidaCoNEXT2017,SSchuppenSec2018}. Our prior work in \cite{MLyuPAM2019} is among the few to analyze DNS traffic in enterprise networks with a view to identifying and monitoring DNS assets to facilitate better security management.

This work is the first to develop data-driven methods to automatically map and monitor DNS assets in an enterprise. Commercial security appliances can consume insights obtained from our methods.
This paper builds upon our prior work in \cite{MLyuPAM2019} (specific extensions and improvements are highlighted in \S\ref{sec:prior}) by drawing new insights into DNS-specific behavior of network assets, developing methods to track cyber-health of assets, and analyzing data of more temporal and spatial coverage collected from two representative enterprise networks. Our contributions are three-fold:

\textbf{First}, in \S\ref{sec:analysis}, we perform a comprehensive forensic analysis of DNS traffic from two large organizations collected over 32 days, comprising nearly a billion queries/responses. We examine their network properties (IPv4/v6, UDP/TCP, etc.), functional properties (unpaired queries/responses, errors, etc.), and service properties (lookup types, record types, etc.). These enable us to build behavioral profiles of how various DNS assets (recursive resolvers, authoritative name servers, and mixed servers) behave.

\textbf{Second}, in \S\ref{sec:class}, we use the behavioral characteristics learned in \S\ref{sec:analysis} to identify key attributes of DNS assets, extract such attributes from network/transport layer header fields without payload inspection, and develop an unsupervised machine learning model using clustering algorithms to dynamically and continuously classify asset types, including recursive resolvers, authoritative name-servers, mixed DNS servers, and regular end-host clients that may or may not be subject to enterprise network address translation (NAT). We apply our method to identify over 100 different DNS assets across the two organizations, and validate our results by cross-checking with IT staff. Our method further identifies assets that were commissioned/decommissioned or changed use during the monitoring period, further validating its utility in dynamically changing environments. 

\textbf{Third}, in \S\ref{sec:health}, we develop data-driven metrics that commercial SIEM\footnote{Security Information and Event Management.} platforms can consume to track the cyber health of DNS assets or identify their anomalous behaviors -- our metrics are inspired by the insights obtained from \S\ref{sec:analysis}. Our methods reveal a prevalence of poor server configurations in both organizations, allowing attackers to exploit them for reflection attacks. Further, our methods are able to identify the organizational DNS assets that are complicit in scans, DDoS, and data exfiltration. We give proposals on how these DNS threats can be mitigated.

Taken together, our contributions help enterprises address their current blind spot in monitoring their DNS assets and the threats they are exposed to. The passive analysis we propose is automated, risk-free, and particularly beneficial to large organizations with numerous assets managed by diverse personnel.

%\vspace{-3mm}
\section{Related Works}\label{sec:prior}
We now discuss related works on DNS traffic analysis (\S\ref{sec:DNSTrafficAnalysis}) and DNS attacks (\S\ref{sec:DNSAttack}) with highlights on the novelty of our work. The details of extension from our previous conference paper \cite{MLyuPAM2019} are explained in \S\ref{sec:ExtensionDetails}.
\subsection{Analysis of DNS Traffic}\label{sec:DNSTrafficAnalysis}
%\vspace{-3mm}
DNS traffic has been analyzed for various purposes, ranging from measuring performance (effect of Time-to-Live of DNS records) \cite{HGaoTNET2016,MMullerIMC2017,MAlmeidaCoNEXT2017} to identifying malicious domains \cite{SHaoIMC2011,SHaoCCS2016,JAhmedIM2019} and the security of DNS \cite{TChungIMC2017,DCMacFarlandPAM2015,DCMacFarlandCN2017,VRijswijk-DeijIMC2014}. 
In this paper we have profiled the pattern of DNS traffic for individual hosts of two enterprise networks to map DNS assets to their function and thereby identify their relative importance and health for efficient monitoring and security. 

DNS data can be collected from different locations (such as from log files of recursive resolvers \cite{HGaoTNET2016,HChoiCN2012} or authoritative name servers) or with different granularity (such as query/response logs or aggregated records). 
Datasets used in \cite{DCMacFarlandPAM2015,DCMacFarlandCN2017,TChungIMC2017} contain DNS traffic for top level domains such as {\myverb{.com}}, and {\myverb{.net}}. 
The work in \cite{DYangATC2020} studies the root cause of query failures by analyzing DNS logs collected from recursive resolvers operated by three Internet service providers.
We collect our data at the edge of an enterprise network, specifically outside the firewall at the point of interconnect with the external Internet.
We note that while using data from resolver logs can provide detailed information about end hosts and their query types/patterns, this approach limits visibility and may not be comprehensive enough to accurately establish patterns related to the assets of the entire network.

Considering studies related to domain names, \cite{SHaoIMC2011} inspects DNS traffic from top-level domain servers to detect abnormal activity and identifies key characteristics of malicious domains in terms of their resource records and lookup patterns, PREDATOR \cite{SHaoCCS2016} derives domain reputation using registration features to enable early detection of potentially malicious DNS domains without capturing traffic, and \cite{WRweyemamuPAM2019} gives practical recommendations for using public domain ranking lists in security research, based on their temporary changes.
%detection of malicious domains
As for detection of such suspicious domain names, \cite{XMATIFS2014} investigates into the coexistence of names in distributed DNS recursive resolvers, \cite{YFuTIFS2017} explores the value of game theory in detecting malicious domain names generated by hidden Markov models and probabilistic context-free grammars, which can bypass legacy detection methods.

\begin{figure}[t!]
	\begin{center}
		{\includegraphics[width=0.46\textwidth, height=0.15\textheight, bb=0 0 1900 800]{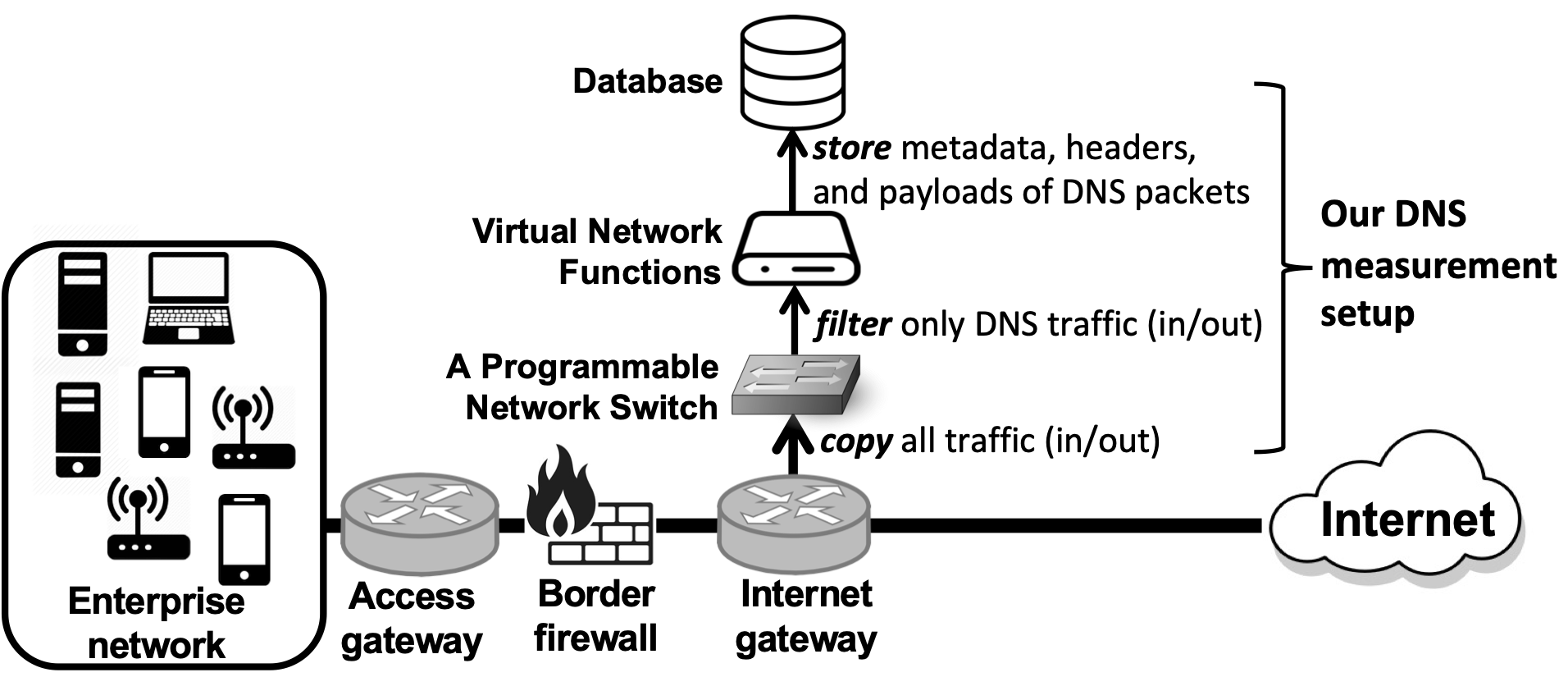}}
		\caption{Our DNS measurement setup.}
		\label{fig:measurementSetup}
	\end{center}
	\vspace{-4mm}
\end{figure}

%our work
\textbf{Our novelty:} Prior works analyzed DNS traffic collected from different vantage points with various objectives. This paper measures traffic at the edge of an enterprise network. We are the first to profile the behavior and health of enterprise hosts by identifying patterns in DNS communications and distribution of various DNS packets. We highlight some interesting observations like when benign query names like {\myverb{google.com}} are misused in cyber-attacks (scans and query-floods) targeting enterprise networks and services.

\subsection{Studies on DNS Attacks}\label{sec:DNSAttack}
From the aspect of DNS attacks, DNSSEC \cite{IETF2018} has been proposed for more than a decade to deprecate information integrity attacks such as cache poisoning \cite{SSonSPCN2010}, however, the authors of \cite{TChungIMC2017} study the adoption of DNSSEC, highlighting that only 1\% of domains have implemented this secure protocol due to difficulties in the registration process and operational challenges -- it is also verified from our results of two enterprises that very little fraction of DNS traffic are mapped to DNSSEC.
The introduction of DNSSEC brings more potentials for volumetric attacks, and some researchers \cite{VRijswijk-DeijIMC2014} have reported that the amplification factor of DNSSEC is quite high (\ie up to 44 to 55) whereas this measure is 6 to 12 for regular DNS servers.
Besides, \cite{DCMacFarlandPAM2015,DCMacFarlandCN2017} focus on authoritative name servers used as reflectors in DNS amplification attacks -- it indicates the potential vulnerabilities of enterprise DNS servers to be mis-used in DDoS attacks.
Work in \cite{MLyuTNSM2021} proposed a hierarchical graph structure with anomaly detection models to identify distributed DNS attackers outside an enterprise network at various levels of aggregation (\eg host, subnet, and AS).

\textbf{Our novelty:} 
Existing works focus on highlighting certain vulnerabilities of the DNS protocol or developing methods for detecting DNS attacks. Our work, instead, systematically profile and track the DNS-related behavior of connected hosts in an enterprise. The system we develop and insights we draw will help IT departments better map their assets, discover potential DNS vulnerabilities, and detect misbehaved (potentially infected) hosts on their network.

\subsection{Extension Details Compared with our Preliminary Work}\label{sec:ExtensionDetails}
This paper is an extension of our work presented as in \cite{MLyuPAM2019}. 
Improvements can be summarized as follows.
We improved our dataset by collecting more inclusive traffic (\ie IPv4 and IPv6) with longer duration (\ie 32 days from 3rd June 2019 to 4th July 2019).
Based on the new dataset, we augmented our forensic analysis by revealing insights of DNS traffic seen at enterprise edge in terms of their `network', `functional' and `service' properties (\S\ref{sec:DNSthreeaspect}); and investigating into composition of normal and malicious DNS packets for each enterprise host (\S\ref{sec:hostanalysis}).
The methodology for DNS asset classification remains unchanged, while we updated corresponding results using the 32-day dataset (\S\ref{sec:class}).
Driven by the security concerns from our analysis, we developed metrics to monitor traffic health of each DNS server, and demonstrate its usage in inferring host anomalies (\S\ref{sec:health}).

\section{Analysis of DNS Traffic from Two Enterprises}\label{sec:analysis}

In this section, we analyze the characteristics of DNS traffic collected from the border of two enterprise networks, a large University campus (\ie the University of New South Wales (UNSW)) and a national research institute (\ie Commonwealth Scientific and Industrial Research Organisation (CSIRO)). 
We start by introducing ur measurement setup is described in \S\ref{sec:measurement}.
In \S\ref{sec:DNSthreeaspect}, we discuss the ``network'', ``functional'', and ``service'' properties of one-week DNS packets collected from both organizations to highlight their normal and abnormal profiles.
We then (in \S\ref{sec:hostanalysis}) focus on the distribution of DNS packets among each enterprise host to reveal their DNS behavioral patterns and unhealthy traffic compositions.

\begin{table*}[t!]
	%	\vspace{-2mm}
	\centering
	\caption{Network properties of DNS packets in our dataset.}
	\label{tab:datasetNetworkUni}
	%	\vspace{-2mm}
	%%\hspace{-4mm}
	\begin{adjustbox}{max width=1\textwidth}    	
		\renewcommand{\arraystretch}{1.5}
		{
			\begin{tabular}{|l|l|l|r|r|r|r|r|r|}
				\hline
				\multicolumn{3}{|c}{} &\multicolumn{3}{|c|}{\textbf{Incoming}}&\multicolumn{3}{|c|}{\textbf{Outgoing}}\\ \cline{4-9}
				\multicolumn{3}{|c|}{} & \textbf{query}       & \textbf{response}  & \textbf{malformed}     & \textbf{query}       & \textbf{response}  & \textbf{malformed} \\ \hline \hline
				\multirow{4}{*}{\rotatebox[origin=c]{90}{\textbf{University}}} 
				& \multirow{2}{*}{\rotatebox[origin=c]{90}{\textbf{IPv4}}} 
				& \textbf{TCP} & \cellcolor{gray!25}$258,315$ & $217,210$  & $\textbf{298,979 (38\%)}$ & $244,633$ & \cellcolor{gray!25}$553,097$&$4,824 (0.3\%)$ \\ \cline{3-9}
				& & \textbf{UDP} & \cellcolor{red!25}$166,492,688$ & \cellcolor{green!25}$181,610,373$ & $\textbf{56,665,050 (14\%)}$ & \cellcolor{green!25}$190,974,279$ & \cellcolor{red!25}$38,321,129$ &$2,158,933(0.9\%)$\\ \cline{2-9}
				& \multirow{2}{*}{\rotatebox[origin=c]{90}{\textbf{IPv6}}} 
				& \textbf{TCP} & $1,223$ & $23,080$  & $5,261(17\%)$ & $25,525$ & $1,203$&$38(0.1\%)$\\ \cline{3-9}
				& & \textbf{UDP} &$10,989,944$ &\cellcolor{blue!25}$53,592,304$ & $200,323(0.3\%)$ &\cellcolor{blue!25}$54,673,191$ & $7,182,025$&$207(0.0006\%)$\\ \cline{1-9} \hline \hline
				\multirow{4}{*}{\rotatebox[origin=c]{90}{\textbf{Rsrch. Ins.}}} 
				& \multirow{2}{*}{\rotatebox[origin=c]{90}{\textbf{IPv4}}} 
				& \textbf{TCP} & $25,829$ & $175,786$  & $18,542(8\%)$ & $200,269$ & $28,421$&$3,375 (1.4\%)$ \\ \cline{3-9}
				& & \textbf{UDP} & \cellcolor{yellow!25}$48,629,262$ & $53,423,998$ & $1,034,531 (1.0\%)$ & $59,638,578$ & \cellcolor{yellow!25}$22,344,154$ &$2,493 (0.003\%)$\\ \cline{2-9}
				& \multirow{2}{*}{\rotatebox[origin=c]{90}{\textbf{IPv6}}} 
				& \textbf{TCP} & $425$ & $11,445$  & $14,394(55\%)$ & $19,708$ & $338$&$274 (1.3\%)$\\ \cline{3-9}
				& & \textbf{UDP} & $5,889,648$ & $14,068,272$ & $82,050 (0.4\%)$ & $16,455,764$ & $6,502,566$&$224(0.0009\%)$\\ \cline{1-9}
		\end{tabular}}
	\end{adjustbox}
	\vspace{-3mm}
\end{table*}
%\vspace{-2mm}

\subsection{Measurement Setup}\label{sec:measurement}

In both organizations, the corresponding IT department provisioned a full mirror (both inbound and outbound) of their Internet traffic (each on a 10 Gbps interface) to our data collection system, shown in Fig.~\ref{fig:measurementSetup}, from their border routers ({\bf outside} of the firewall). 
Note that the outgoing portion of our measured traffic is already filtered by firewall policies, while the incoming portion remains unfiltered (and hence may be relatively polluted, as discussed in \S\ref{sec:NetworkProperty}, indicative of wild Internet traffic. That said, we mainly focus on traffic corresponding to successful two-way communications to compute attributes needed for asset classification (\S\ref{sec:class}) and metrics needed for asset health tracking (\S\ref{sec:health}). 
Appropriate ethics clearances for this study are granted\footnote{UNSW Human Research Ethics Advisory Panel approval number HC17499, and CSIRO Data61 Ethics approval number 115/17.}. 
We extracted DNS packets from the Internet traffic stream of each enterprise in real-time by configuring rules for incoming/outgoing IPv4 and IPv6 TCP/UDP packets for port 53 on a programmable network switch. In this paper, we focus on DNS traffic via its typically assigned port 53, while it is worth noting that a tiny fraction of DNS lookups might be carried by TLS \cite{rfcDNSTLS2018} and HTTPS \cite{rfcDNSHTTPS2018} that are beyond the scope of this paper.
The mirrored DNS traffic was processed by a virtual network function running on a generic server (with DPDK \cite{DPDK2020}) which parses headers (network, transport, application) and payload of each DNS packet, and stores them into our database.  
The study in this paper considers the data collected over a month period of 3 June to 4 July 2019 (\ie beginning of an academic term in the university).
In this section, we focus on the analytic results for 1 week worth of DNS traffic capture from both organizations during 3 June to 9 June 2019.

\subsection{Understanding DNS Traffic at Enterprise Network Border}\label{sec:DNSthreeaspect}
We begin by examining ``network'', ``functional'', and ``service'' properties of DNS packets, which provide answers for the following three questions related to DNS traffic profiles of an organization. 
How does each DNS packet get carried at network-level (\S\ref{sec:NetworkProperty})?
Is each DNS packet with correct or error functionality (\S\ref{sec:FunctionalProperty})?
What is the service type of each DNS packet (\S\ref{sec:ServiceProperty})?

\subsubsection{Network Property}\label{sec:NetworkProperty}
DNS packets can be carried by either TCP or UDP at the transport layer via IPv4 or IPv6 protocols. 
Table~\ref{tab:datasetNetworkUni} summarizes the composition of DNS packets (in our dataset) by their network properties.

\textbf{IPv4 versus IPv6:} Unsurprisingly, the majority of DNS packets are carried by IPv4 protocol, and it is clear that the adoption of IPv6 in DNS communications has become non-negligible in both organizations.
We found that $21.03\%$ and $21.84\%$ of outgoing DNS packets in the university and research networks, respectively, are IPv6, while this measure for incoming DNS traffic of the two organizations is $13.78\%$ and $16.26\%$.

\textbf{TCP versus UDP:} Considering the transport layer, DNS over UDP seems to be default for enterprise hosts, accounting for more than $99$\% of outgoing and incoming packets in both organizations, while DNS over TCP is still staying minority (less than $0.3$\%). We note that DNS occasionally uses TCP when the size of the request or the response is greater than a single packet such as with responses that have many records or many IPv6 responses.

\textbf{Queries versus Responses:} Focusing on the correlation between DNS queries and responses, we highlight four pairs of query/response in Table~\ref{tab:datasetNetworkUni}, as examples -- each pair is color-coded for identification. 
It can be seen that the number of outgoing queries is slightly higher than the number of incoming responses, suggesting unanswered  DNS lookups made by enterprise hosts (green and purple pairs in Table~\ref{tab:datasetNetworkUni}). 
We also observe that count of incoming queries over IPv4 UDP is more than double the count of outgoing responses in both organizations (\eg red and yellow pairs in Table~\ref{tab:datasetNetworkUni}), highlighting the prevalence of DNS scans and floods on enterprise networks. However, this is not substantiated in IPv6 packets. Lastly, we note that in the university network, the count of outgoing TCP-based responses over IPv4 is more than double the number of their corresponding queries (the gray pair in Table~\ref{tab:datasetNetworkUni}). Manually analyzing those unmatched outgoing TCP responses revealed that they are ACK packets sourced from TCP/53 (probably crafted to look like DNS responses to bypass firewalls). We found that those TCP ACK packets are generated by 640 enterprise IP addresses (that are end-hosts) -- none of them are classified as authoritative name server or recursive resolver later in \S\ref{sec:class}.
Such behaviors are often seen in TCP-based malicious TCP activities such as ACK-based host scans \cite{ACKPING} or ACK flooding attacks \cite{ACKFlood}.

\textbf{Malformed Packets:} We found that $9.8$\% and $0.5$\% of total DNS packets in the university and research institute datasets, respectively, are malformed. These packets cannot be correctly parsed as their DNS header information mismatched the payload content (\eg the number of resource records indicated in the header is not consistent with the actual content in the payload). There are various reasons for having malformed DNS packets mentioned in \cite{MBykovaSSST2001,MBykovaIMC2002,JMirkovicCCR2004} such as malicious traffic crafted by attackers and packet truncation or distortion during transmission. It can be seen that there are more malformed incoming packets compared to outgoing packets, as highlighted by percentage values (computed per each row per direction) under malformed columns in Table~\ref{tab:datasetNetworkUni}. We note that all malformed packets result in no response (\ie probably they get filtered by the border firewall or dropped by the destination host). %, thus, failing to trigger effective DNS lookups.}

Another observation is that malformed DNS packets are more likely carried over TCP. For example, an inbound packet over IPv4 TCP in the university network is malformed with a probability of $38$\%, while that is $14$\% over IPv4 UDP (bold text in Table~\ref{tab:datasetNetworkUni}). Besides, when comparing the two organizations, we observed that the university network sends more malformed packets in total fraction than the research institution, particularly for outbound IPv4 UDP packets ($0.9$\% versus $0.003$\% for the university and the research institute, respectively). It indicates frequent malicious activities originated from university hosts, as the university network is open and less restricted, while the research institute does not allow BYOT (bring-you-own-technology) devices and has strict enforcement for network security.

\begin{table}[t!]
	\caption{Functional properties of DNS packets.}
	\label{tab:datasetFunction}
	%	\vspace{-2mm}
	\hspace{-3mm}
	\centering
	\begin{adjustbox}{max width=0.5\textwidth}    	
		\renewcommand{\arraystretch}{1.2}
		{
			\begin{tabular}{|c|r|r|r|r|r|}
				\hline
				\multicolumn{2}{|c}{} &\multicolumn{2}{|c|}{\textbf{Incoming}}&\multicolumn{2}{|c|}{\textbf{Outgoing}}\\ \cline{3-6}
				\multicolumn{2}{|c|}{} & \textbf{IPv4}       & \textbf{IPv6}       & \textbf{IPv4}       & \textbf{IPv6}   \\ \hline \hline
				\multirow{8}{*}{\rotatebox[origin=c]{90}{\textbf{University}}} 
				& \textbf{Unanswered qry.} & \cellcolor{red!25}$130,683,135$ & \cellcolor{red!25}$3,813,677$  & \cellcolor{red!25}$11,431,123$ & \cellcolor{red!25}$1,105,818$ \\ \cline{2-6}
				& \textbf{Unsolicited resp.} & $2,039,794$ & $22,486$  & $2,806,358$ & $5,738$ \\ \cline{2-6}
				& \textbf{NameError pairs} & $7,493,599$ & $1,885,742$  & \cellcolor{yellow!25}$5,164,713$ & \cellcolor{yellow!25}$1,532,410$ \\ \cline{2-6}
				& \textbf{Serv.Failure pairs} & $3,897,549$ & $34,643$  & $1,363,391$ & $112,618$ \\ \cline{2-6}
				& \textbf{Qry.Refused pairs} & \cellcolor{yellow!25}$24,820,580$ & \cellcolor{yellow!25}$16,409,541$  & $2,102,724$ & $26,130$ \\ \cline{2-6}
				& \textbf{OtherError pairs} & $113,291$ & $90$  & $794$ & $0$ \\ \cline{2-6}
				& \textbf{Non-enterprise pairs} & $9,153,748$ & $252,860$  & $178,234,417$ & $53,096,280$ \\ \cline{2-6}
				& \textbf{Enterprise pairs} & $26,914,120$ & $6,924,630$ & $1,553,372$ & $496,618$ \\ \cline{1-6} \hline \hline
				\multirow{8}{*}{\rotatebox[origin=c]{90}{\textbf{Research Institute}}} 
				& \textbf{Unanswered qry.} & \cellcolor{red!25}$29,159,158$ & \cellcolor{red!25}$182,886$  & \cellcolor{red!25}$9,912,604$ & \cellcolor{red!25}$2,843,892$ \\ \cline{2-6}
				& \textbf{Unsolicited resp.} & $3,673,541$ & $448,137$  & $2,876,642$ & $795,717$ \\ \cline{2-6}
				& \textbf{NameError pairs} & \cellcolor{yellow!25}$2,730,158$ & \cellcolor{yellow!25}$974,480$  & \cellcolor{yellow!25}$3,775,508$ & \cellcolor{yellow!25}$1,011,591$ \\ \cline{2-6}
				& \textbf{Serv.Failure pairs} & $248,275$ & $19,715$  & $2,389,070$ & $5,601$ \\ \cline{2-6}
				& \textbf{Qry.Refused pairs} & $1,390,138$ & $245,390$  & $781,259$ & $133,599$ \\ \cline{2-6}
				& \textbf{OtherError pairs} & $17,061$ & $50$  & $621$ & $230$ \\ \cline{2-6}
				& \textbf{Non-enterprise pairs} & $2,071,310$ & $558,205$  & $48,754,035$ &  $13,275,249$ \\ \cline{2-6}
				& \textbf{Enterprise pairs} & $17,424,623$ & $5,148,982$ & $1,172,208$ & $356,331$ \\ \cline{1-6}		
		\end{tabular}}
	\end{adjustbox}
	\vspace{-5mm}
\end{table}

\subsubsection{Functional Property}\label{sec:FunctionalProperty}

In terms of functional property, we categorize DNS packets into three clusters: (a) unpaired packets (\ie queries with no reply or responses without a corresponding query), (b) DNS lookups with a reply containing errors, and (c) successful DNS lookups.

\textbf{Unpaired Packets:}
This category is captured by two rows labeled as ``\textit{unanswered qry.''} and ``\textit{unsolicted resp.}'' in Table \ref{tab:datasetFunction}. 
Unanswered queries (highlighted by red cells in Table~\ref{tab:datasetFunction}), carried over both IPv4 and IPv6, contribute to a large fraction of total incoming DNS packets -- $40.4$\% and $30.8$\% in the university and research institute, respectively -- this is due to frequent DNS scans and query floods targeting enterprise DNS infrastructure. 
On the other hand, unanswered outgoing queries only account for a relatively smaller fraction in each organization (\ie $2.4$\% and $7.9$\%).
Moving to unsolicited responses, their fraction in both inbound and outbound traffic are quite similar. This is mainly because of packet drop during transmission, misconfiguration of external DNS servers or DNS-based reflection attacks from/to the enterprises \cite{HGaoSIGCOMM2013},\cite{MAnagnostopoulosNOMS2018}.

\textbf{DNS Lookups with Error Reply:}
Now we focus on the DNS lookups with error replies, classified by their {\myverb{errorCode}} in the header of response packets. 
Top three popular error types in both enterprises are listed as {\myverb{NameError}}, {\myverb{ServerFailure}} and {\myverb{QueryRefused}} in Table~\ref{tab:datasetFunction} -- all other errors are grouped under {\myverb{OtherError}}.
{\myverb{NameError}}, also known as {\myverb{Non-Existence Domain}} (will be referred as {\myverb{NXDOMAIN}} for simplicity in the rest of this paper \eg \S\ref{sec:health}), is caused if the requested domain name is incorrect (does not exist). It might be because of typos made by legitimate users or malicious queries (\eg command-and-control and data exfiltration) sent by malware-infected hosts \cite{MAntonakakisSec2012,SSchuppenSec2018}. 
{\myverb{ServerFailure}} and {\myverb{QueryRefused}} indicate that target DNS servers are not able to provide a resolved answer for various reasons such as zone restrictions or incorrect query formats. A frequent occurrence of those errors could indicate improper configurations on hosts/servers, or DNS attacks.
As highlighted by yellow cells in Table~\ref{tab:datasetFunction}, for the university network, {\myverb{QueryRefused}} and {\myverb{NameError}} are the most popular error types of incoming and outgoing lookups, respectively; while 
{\myverb{NameError}} dominates the error types of both incoming and outgoing lookups in the research institute.

\begin{table}[t!]
	%	\vspace{-3mm}
	\caption{Service properties of DNS packets in our dataset.}
	\label{tab:datasetServiceType}
	%	\vspace{-2mm}
	\hspace{-2mm}
	%\centering
	\begin{adjustbox}{max width=0.49\textwidth}    	
		\renewcommand{\arraystretch}{1.2}
		{
			\begin{tabular}{|c|r|r|r|r|r|}
				\hline
				\multicolumn{2}{|c}{} &\multicolumn{2}{|c|}{\textbf{Incoming}}&\multicolumn{2}{|c|}{\textbf{Outgoing}}\\ \cline{3-6}
				\multicolumn{2}{|c|}{} & \textbf{IPv4}       & \textbf{IPv6}       & \textbf{IPv4}       & \textbf{IPv6}   \\ \hline \hline
				\multirow{11}{*}{\rotatebox[origin=c]{90}{\textbf{University}}} 
				& \textbf{A pairs} & \cellcolor{blue!25}$19,986,211$ & \cellcolor{blue!25}$3,692,671$  & \cellcolor{blue!25}$111,896,351$ & \cellcolor{blue!25}$29,538,537$ \\ \cline{2-6}
				& \textbf{AAAA pairs} & \cellcolor{blue!25}$7,782,897$ & \cellcolor{blue!25}$2,014,541$ & \cellcolor{blue!25}$32,223,426$ & \cellcolor{blue!25}$6,615,980$ \\ \cline{2-6}
				& \textbf{PTR pairs} & \cellcolor{blue!25}$2,927,101$ & \cellcolor{blue!25}$594,635$  & \cellcolor{blue!25}$24,749,068$ & \cellcolor{blue!25}$15,775,549$ \\ \cline{2-6}
				& \textbf{MX pairs} & $1,413,019$ & $210,452$  & $831,571$ & $192,365$ \\ \cline{2-6}
				& \textbf{SPF pairs} & $43,943$ & $8,600$  &\cellcolor{yellow!25} $109$ & \cellcolor{yellow!25}$28$ \\ \cline{2-6}
				& \textbf{TXT pairs} & $723,796$ & $64,690$  & $4,415,435$ & $659,723$ \\ \cline{2-6}
				& \textbf{CNAME pairs} & $79,235$ & $23,408$  & $11,708$ & $1,693$ \\ \cline{2-6}
				& \textbf{SRV pairs} & $599,022$ & $197,023$  & $2,678,513$ & $103,494$ \\ \cline{2-6}
				& \textbf{SOA pairs} & $220,711$ & $88,752$  & $714,524$ & $299,316$ \\ \cline{2-6}
				& \textbf{NS pairs} & $1,057,700$ & $223,808$  & $727,438$ & $358,427$ \\ \cline{2-6}
				& \textbf{ANY pairs} & \cellcolor{red!25}\textbf{1,205,822} & $46,315$  &  \cellcolor{red!25}$114,584$ &  \cellcolor{red!25}$9,592$ \\ \cline{2-6}
				& \textbf{Other pairs} & $21,990$ & $10,305$ & $1,209,553$ & $3,754$ \\ \cline{1-6} \hline \hline
				\multirow{11}{*}{\rotatebox[origin=c]{90}{\textbf{Research Institute}}} 
				& \textbf{A pairs} &  \cellcolor{blue!25}$7,664,442$ &  \cellcolor{blue!25}$1,585,811$  & \cellcolor{blue!25}$21,571,867$ &  \cellcolor{blue!25}$6,174,823$ \\ \cline{2-6}
				& \textbf{AAAA pairs} &\cellcolor{blue!25}$2,287,039$ & \cellcolor{blue!25}$755,134$ & \cellcolor{blue!25}$23,774,650$ & \cellcolor{blue!25}$6,107,818$ \\ \cline{2-6}
				& \textbf{PTR pairs} &\cellcolor{blue!25}$7,677,620$ & \cellcolor{blue!25}$2,998,552$  & \cellcolor{blue!25}$2,040,424$ & \cellcolor{blue!25}$599,030$ \\ \cline{2-6}
				& \textbf{MX pairs} & $441,075$ & $117,015$  & $301,651$ & $84,904$ \\ \cline{2-6}
				& \textbf{SPF pairs} & $3,782$ & $662$  & $15,974$ & $3,984$ \\ \cline{2-6}
				& \textbf{TXT pairs} & $120,308$ & $13,198$  & $1,099,786$ & $342,250$ \\ \cline{2-6}
				& \textbf{CNAME pairs} & $43,399$ & $5,556$  & $19,933$ & $116$ \\ \cline{2-6}
				& \textbf{SRV pairs} & $230,046$ & $33,483$  & $364,039$ & $79,366$ \\ \cline{2-6}
				& \textbf{SOA pairs} & $222,796$ & $40,515$  &\cellcolor{yellow!25} $0$ &\cellcolor{yellow!25} $0$ \\ \cline{2-6}
				& \textbf{NS pairs} & $683,641$ & $132,410$  & $532,757$ & $179,340$ \\ \cline{2-6}
				& \textbf{ANY pairs} & $101,867$ & $23,012$  & \cellcolor{red!25}$830$ & \cellcolor{red!25}$447$ \\ \cline{2-6}
				& \textbf{Other pairs} & $18,054$ & $988$ & $1,512$ & $165$ \\ \cline{1-6}	
		\end{tabular}}
	\end{adjustbox}
	\vspace{-4mm}
\end{table}

\textbf{Successful DNS Lookups:}
Given a successful pair of DNS packets (\ie DNS query-response pair with {\myverb{NoError}} flag), their requested domain names can be classified as either relevant (\ie belong to services provided by the organization) or irrelevant.
For organizations running their authoritative name servers for their domain names, we expect that all inbound DNS lookups are relevant to the enterprise. However, a non-negligible portion of inbound non-error DNS lookups asks for irrelevant domain names in both enterprises ($21.7$\% and $10.4$\%, respectively). They are likely to be malicious DNS queries involved in scans or floods; furthermore, improperly configured DNS servers in the enterprise resolved those irrelevant questions.
Outbound DNS lookups for two organizations contain a tiny portion of questions for the enterprise services ($0.8$\% and $2.4$\%). As verified in our dataset, the top destinations of those queries are public recursive resolvers such as {\myverb{8.8.8.8}} and  {\myverb{8.8.4.4}} operated by Google. It shows some hosts (for some reason) bypassed the local DNS caches, operational within each enterprise, by choosing public resolvers for their DNS server.

\subsubsection{Service Property}\label{sec:ServiceProperty}

Successful DNS lookups are asking for various types of services, such as IPv4 address (A type), IPv6 address (AAAA type), and reverse lookup for domain names (PTR type). 
We now focus on the question type specified in the query header of each successful lookup. Statistics for success inbound/outbound lookup pairs are shown in Table \ref{tab:datasetServiceType}.

\textbf{Successful Lookups}: We start with successful lookup types in both networks. 
As highlighted by blue cells in Table \ref{tab:datasetServiceType}, as expected, we observe that requests for IPv4/IPv6 addresses and reverse lookups for domain names are most common types of both inbound and outbound DNS traffic in both organizations.
Besides, non-negligible amounts of email-related (\ie {\myverb{MX}} and {\myverb{SPF}}), text exchange ({\myverb{TXT}}), authoritative name service-related (\ie {\myverb{CNAME}}, {\myverb{NS}} and {\myverb{SOA}}), and service location ({\myverb{SRV}}) lookups are observed in both organizations, indicating the popularity of their corresponding services.
Note that {\myverb{CNAME}}, {\myverb{NS}} and {\myverb{SOA}} are generated by authoritative name servers to indicate canonical names, name servers, and authoritative domains, respectively. On the University network, those outgoing responses are generated by 25 internal hosts that all will be classified as authoritative name servers later in \S\ref{sec:class} -- two main servers contributed to more than 95\% of those packets.
Apart from these top contributing types consistently seen in both networks, we observe some different service profiles across the two organizations. For example, as highlighted by yellow cells in Table~\ref{tab:datasetServiceType}, no outbound DNS lookup is found for {\myverb{SOA}} (that asking for authoritative information of a zone) in the research institute, while a few outbound lookups for {\myverb{SPF}} (requesting authorized email servers of a domain) are seen in the university network.

\vspace{3mm}
\textbf{Deprecated Services:} We discovered some obsolete types of DNS lookups. For example, the deprecation of {\myverb{ANY}} type requests was announced by the networking community \cite{Cloudflare2019,rfcDNSANYStop2021} for their frequent misuses in reflection attacks and rare legitimate uses (\eg resolver debugging and testing) today. However, we found a large number of {\myverb{ANY}} type DNS lookups (inbound traffic) in both organizations, particularly for the university network (marked as bold red text). 
Focusing on the outbound traffic, university hosts sent out many such types of deprecated requests. In contrast, hosts in the research institute rarely had such activities (relevant cells are marked as red). Besides, many {\myverb{A6}} (deprecated version of lookups for IPv6 address) and {\myverb{NAPTR}} (mapping domain names to host URLs) are found in outbound requests in the university network, respectively contributing to $0.15\%$ and $0.31\%$ of the total count of outbound queries.
%total univ out query = 245,917,628
%A6: 0.15% - 356867 NAPTR:  0.31%-773491

\textbf{Adoption of DNSSEC:} We now look at statistics of DNS lookups related to DNSSEC in both organizations. DNSSEC \cite{IETF2018} has been proposed for more than a decade to strengthen information integrity of DNS data, prior measurement studies \cite{TChungIMC2017} on domain registrars resulted that the adoption of such extension is still in early stage. Fairly similar observations were made in both networks, as there are $0.005\%$ inbound lookups and $0.1\%$ outbound lookups are associated with DNSSEC services in the university campus, including {\myverb{DNSKEY}},  {\myverb{DS}}, {\myverb{RRSIG}}, {\myverb{NSEC}}, {\myverb{NSEC3}} and {\myverb{DLV}}. The fraction value of such inbound and outbound lookups for the research institute are $0.005\%$ and $0.2\%$, respectively. 
In addition, outbound lookup fractions in both organizations are much larger than those of inbound ones, indicating that their authoritative name servers do not support DNSSEC.

\begin{figure}[t!]
	%	\vspace{-3mm}
	\begin{center}
		\mbox{
			\hspace{-5mm}
			\subfigure[Outgoing DNS queries.]{
				{\includegraphics[width=0.25\textwidth]{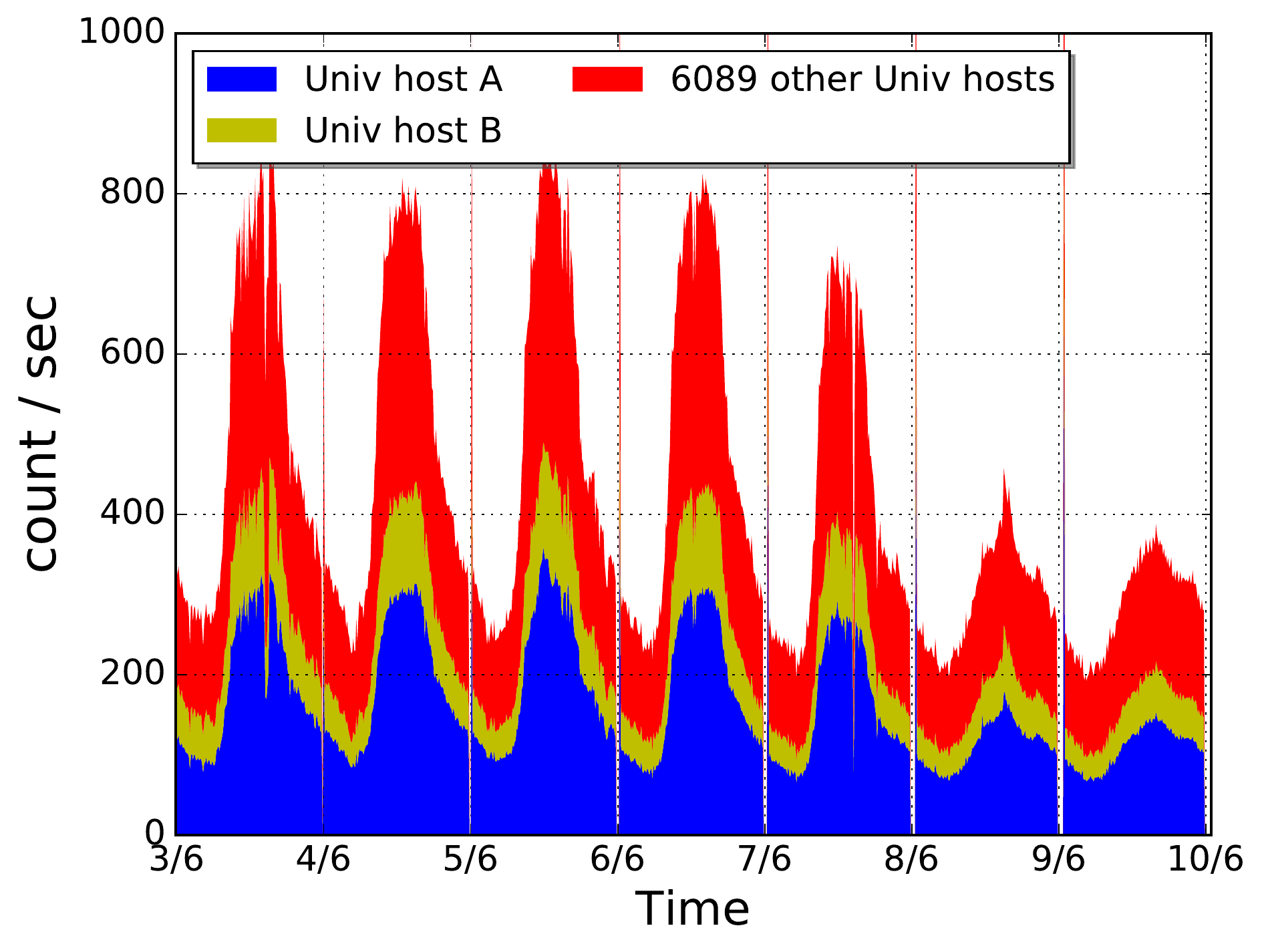}}\quad
				\label{fig:outgoingqueriesUNSW}
			}
			\hspace{-7mm}
			\subfigure[Incoming DNS responses.]{
				{\includegraphics[width=0.25\textwidth]{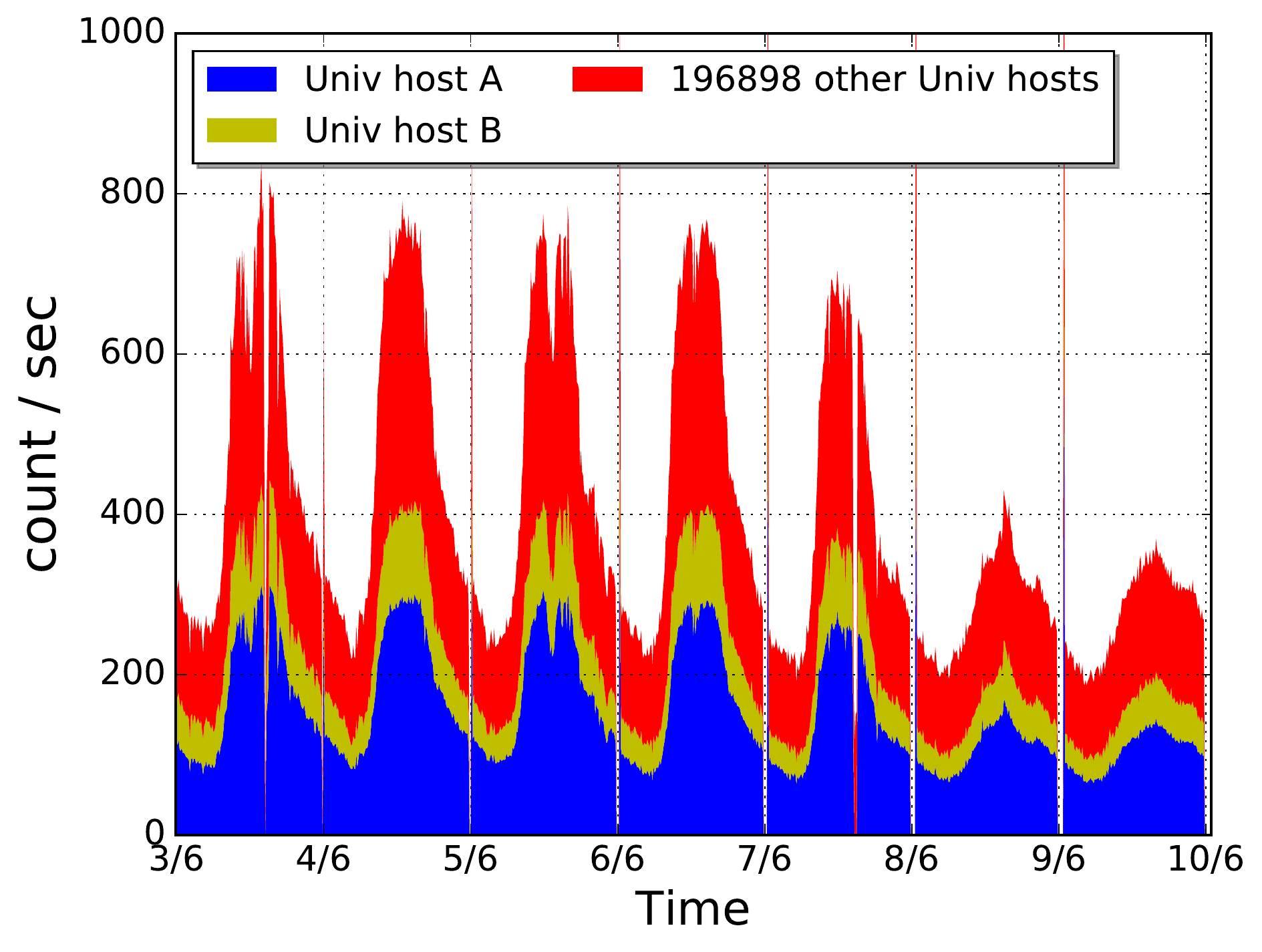}}\quad
				\label{fig:incomingresponsesUNSW}
			}
			
		}
		%		\vspace{-3mm}
		\caption{University campus: outgoing queries and incoming responses, measured during 3 June to 9 June 2019.}
		%	\vspace{-5mm}
		\label{fig:outQryUNSW}
	\end{center}
	\vspace{-4mm}
\end{figure}

\subsection{Profiling DNS Behaviors of Enterprise Hosts}\label{sec:hostanalysis}

Enterprises typically operate two types of DNS servers: (a) {\bf recursive resolvers} are those that act on behalf of end-hosts to resolve the network address of a domain name and return the answer to the requesting end-host (recursive resolvers commonly keep a copy of positive responses in a local cache for time-to-live of the record to reduce frequent recursion), and (b) {\bf authoritative servers} of a domain/zone are those that receive queries from anywhere on the Internet for the network address of a sub-domain within the zone for which they are authoritative (\eg {\myverb{organizationXYZ.net}}).

In order to better understand the DNS behavior of various hosts (and their role) inside an enterprise network, we divide the DNS dataset into two categories: (a) DNS queries from enterprise hosts that leave the network towards a server on the Internet along with DNS responses that enter the network (\S\ref{sec:OutQryInResp}), (b) DNS queries from external hosts that enter the network towards an enterprise host along with DNS responses that leave the network (\S\ref{sec:InQryOutResp}).

This analysis helps us identify important attributes related to host DNS behavior, characterizing its type/function, including authoritative name server, recursive resolver, or end-host inside the enterprise that may not always be fully visible to the network operators. This also enables us to capture the normal pattern of DNS activity for various hosts and identify the abnormal traffic status of DNS infrastructures. 

\begin{figure}[t!]
	%	\vspace{-3mm}
	\begin{center}
		\mbox{
			\hspace{-5mm}
			\subfigure[CCDF: \# Unwanted DNS pkts per host.]{
				{\includegraphics[width=0.25\textwidth]{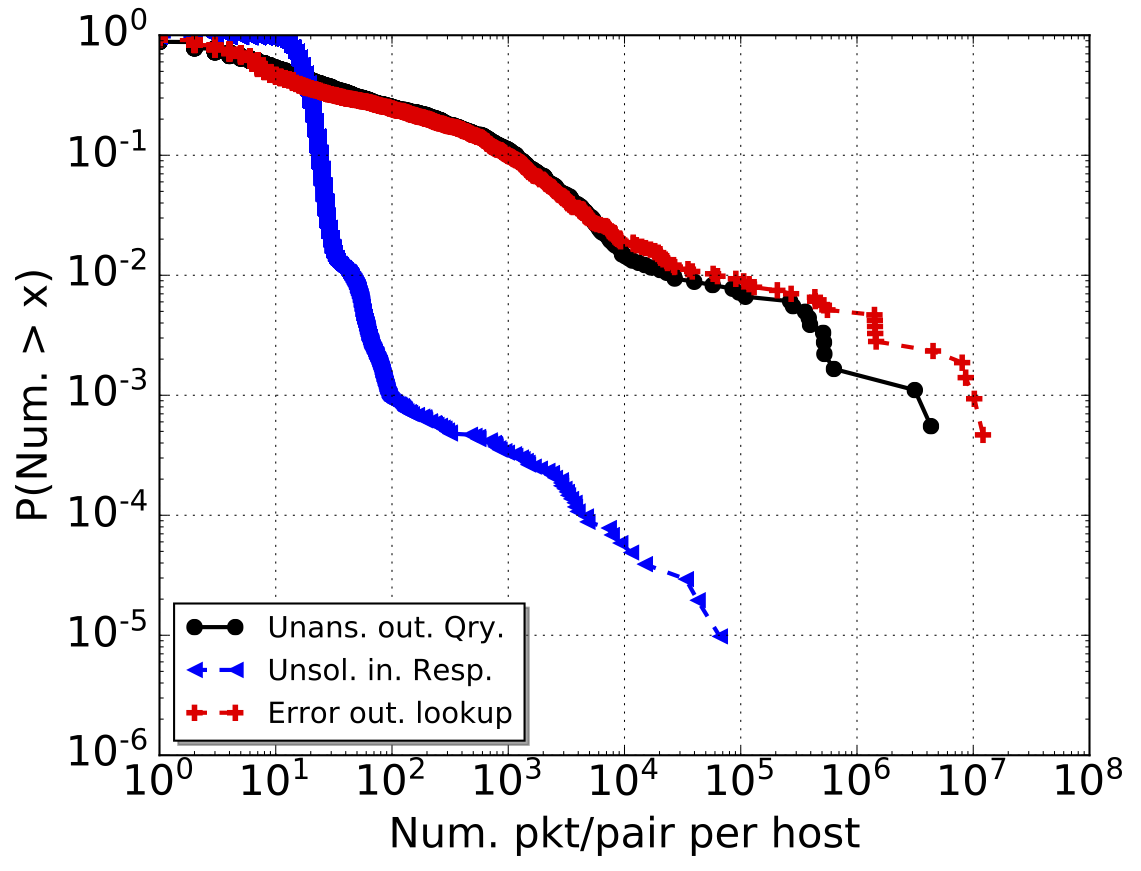}}\quad
				\label{fig:ccdfUansweredQryOutUNSW}
			}
			\hspace{-6mm}
			\subfigure[Traffic composition per host.]{
				{\includegraphics[width=0.25\textwidth]{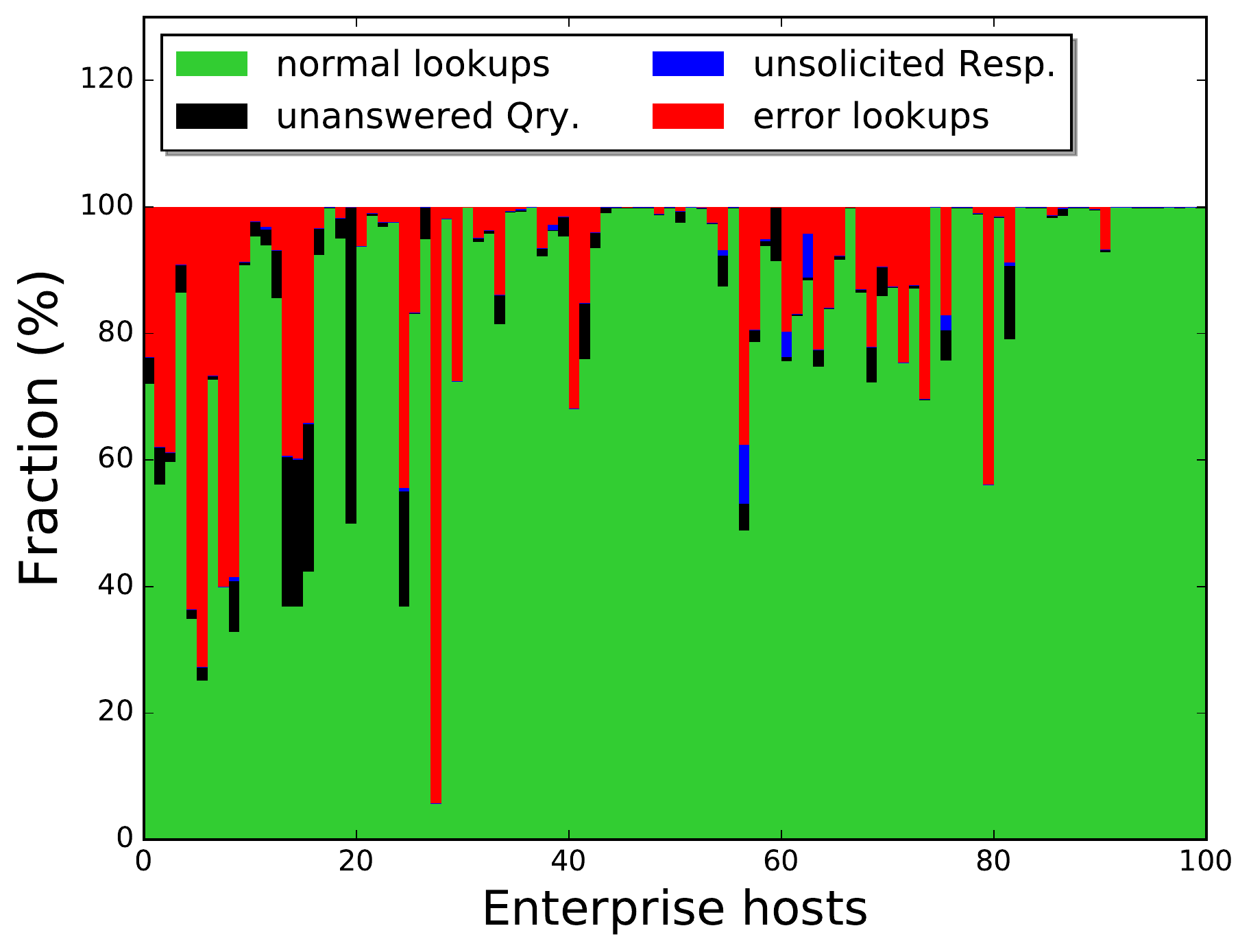}}\quad
				\label{fig:ccdfUnsolicitRespInUNSW}
			}
		}
		%	\vspace{-3mm}
		\caption{University campus: (a) CCDF of \# unwanted (outgoing queries and incoming responses) DNS packets and (b) their total fractions per host, measured during 3 June to 9 June 2019.}
		%		\vspace{-5mm}
		\label{fig:ccdfUnsolicitRespInUnansweredQryOutUNSW}
	\end{center}
	\vspace{-8mm}
\end{figure}

\subsubsection{Outgoing Queries \& Incoming Responses}\label{sec:OutQryInResp}

Fig.~\ref{fig:outQryUNSW} shows a time trace of DNS outgoing queries and incoming responses for the university campus\footnote{We omit results for the research institute in this section, as fairly similar observations were made.}, with granularity over 10-minute intervals on a typical semester week. 

The university network handles on average 417 outgoing queries and 408 incoming responses per second. As discussed in Table \ref{tab:datasetNetworkUni}, $4.9\%$ of outgoing queries are ``unanswered'' (\ie $12.5M$ out of $256.2M)$ during the week. 
And $2.06\%$ of incoming responses to the university campus network (\ie $2.1M$ out of $99.9M$) are ``unsolicited'' on the same day.

\textbf{Query Per Host:}
We now consider individual hosts in each enterprise. Unsurprisingly, the majority of outgoing DNS queries are generated by only two hosts, A and B, in the network, \ie $66.8\%$ of the total in the university campus (shown by blue and yellow shades in Figures~\ref{fig:outgoingqueriesUNSW}). These hosts are also the primary recipients of incoming DNS responses from the Internet. We have verified with the IT department of the enterprise that both hosts are primary recursive resolvers of this organization.
In addition to these recursive resolvers, we observe a number of hosts shown by red shades in Fig.~\ref{fig:outgoingqueriesUNSW} that generate DNS queries outside of the enterprise network. The 6,089 other University hosts in Fig.~\ref{fig:outgoingqueriesUNSW} are either: end-hosts configured by public DNS resolvers that make direct queries out of the enterprise network, or secondary recursive servers operating in smaller sub-networks at the department level. We found that 301 of these 6,089 University hosts actively send queries (at least once every hour) over the day and contact more than 10 Internet-based DNS servers (resolvers or name-servers). These 301 hosts display the behavior of recursive resolvers but with fairly low throughput; thus, we deem them secondary resolvers. The remaining 5,788 hosts are only active for a limited interval (\ie between 5 min to 10 hours) and contact a small number of public resolvers (\eg 8.8.8.8 or 8.8.4.4 of Google) over the day.

\textbf{Response Per Host:}
Considering incoming responses  in Fig.~\ref{fig:incomingresponsesUNSW} for the university network, a larger number of ``other'' hosts in the organization are observed -- approximately 196K IP addresses corresponding to the three subnets of size /16 owned by the university. Most of these ``other'' hosts (\ie $97\%$) are the destinations of unsolicited responses, which indicates that either misconfiguration of external DNS servers, or the university network is suffering from DNS reflections.

\textbf{Unwanted DNS Packets Per Host:}
To better understand these potentially abnormal unanswered outgoing queries, unsolicited incoming responses, and error outgoing DNS lookups, we analyze their distribution among hosts in the two enterprises. 

Fig.~\ref{fig:ccdfUansweredQryOutUNSW} shows the CCDF plot of the distributions per host for the university campus. 
All enterprise IP addresses in our dataset received unsolicited responses, and it is clear from the blue line that 99.9\% of them are associated with 10 to 100 such packets -- they did not have any outbound queries over the week. We observe that the hosts that have sent outbound queries to the public Internet received more unsolicited responses than those hosts that have never sent any DNS lookup.
Outbound unanswered queries and error lookups are more concentrated on a small fraction of hosts, as shown in the tail of black and red lines. 2,140 and 1,812 (out of 6,091) hosts sent unanswered or error lookups -- possibly due to packet drop during forwarding, typo-error of domain names, or malicious activities such as generating scans and DoS attacks.

Unsurprisingly, the primary recursive resolvers in both organizations are top sources and targets. In the University campus, hosts A and B respectively are the sources $4M$ ($33\%$) and $3M$ ($25\%$) unanswered queries, $12M$ ($22\%$) and $10M$ ($18\%$) error lookups, and are the destinations of $66K$ ($3\%$) and $42K$ ($2\%$) unsolicited responses.

\begin{figure}[t!]
	%	\vspace{-3mm}
	\begin{center}
		\mbox{
			\hspace{-5mm}
			\subfigure[Incoming DNS queries.]{
				{\includegraphics[width=0.25\textwidth]{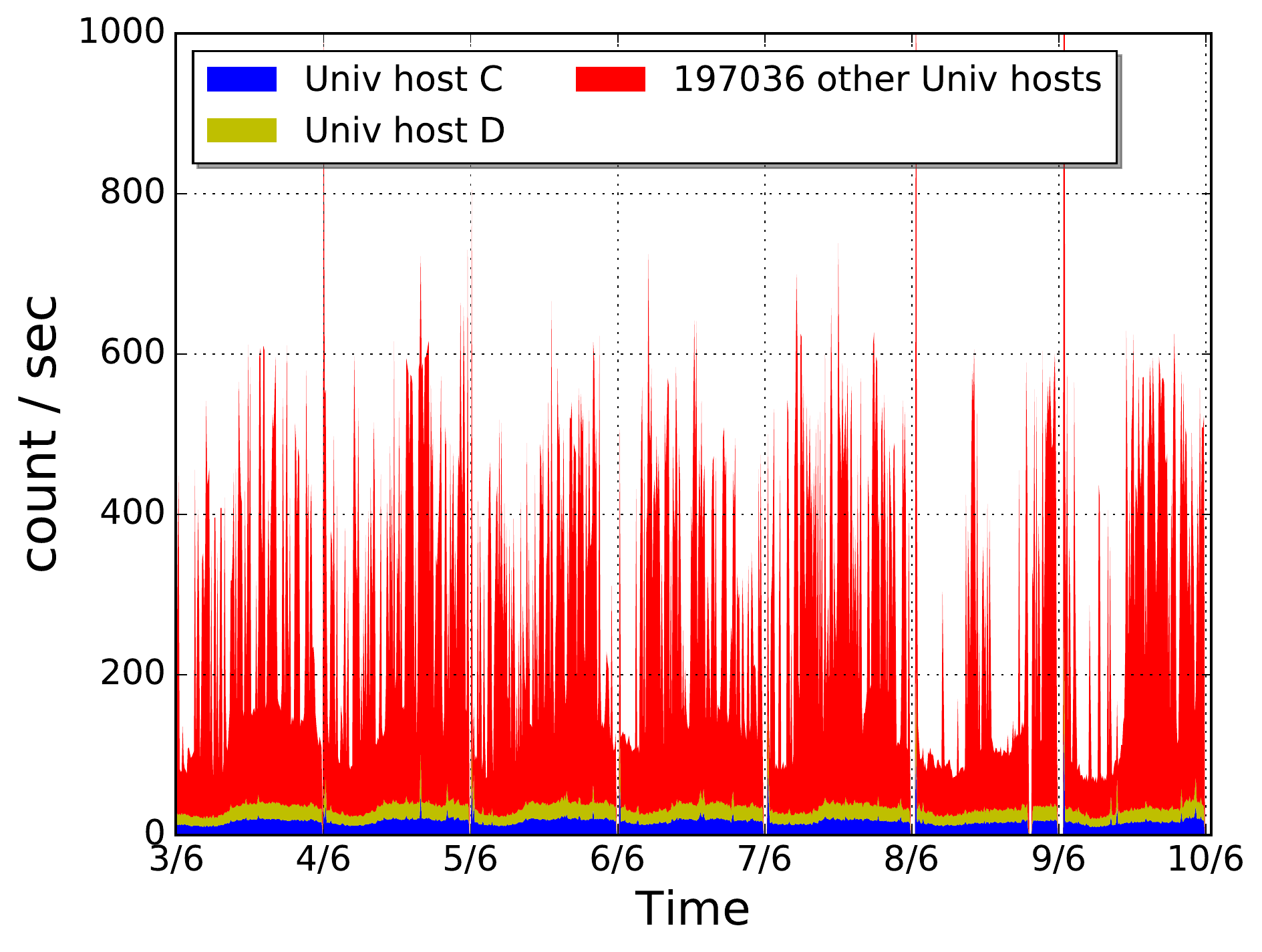}}\quad
				\label{fig:incomingqueriesUNSW}
			}
			\hspace{-7mm}
			\subfigure[Outgoing DNS responses.]{
				{\includegraphics[width=0.25\textwidth]{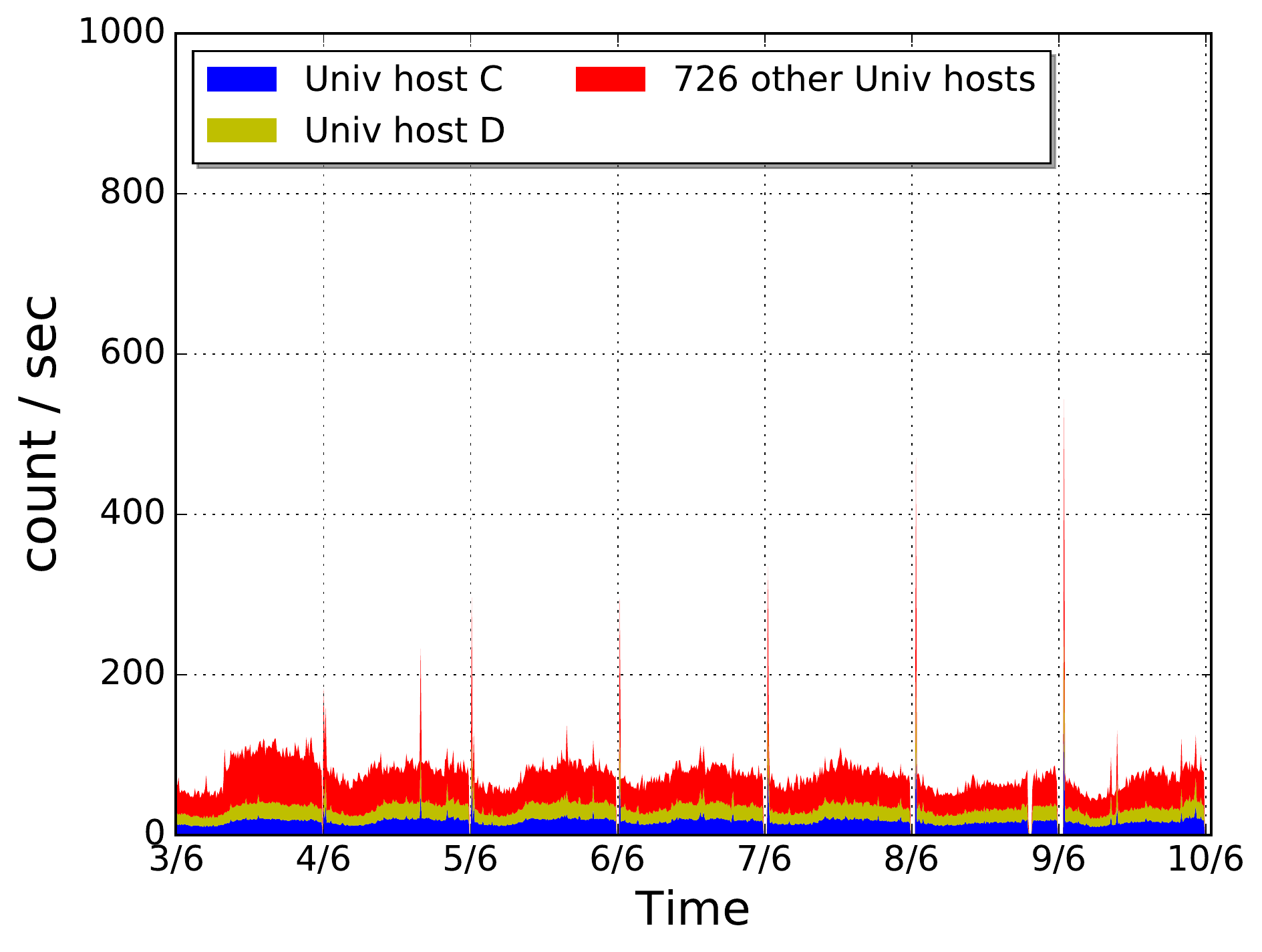}}\quad
				\label{fig:outgoingresponsesUNSW}
			}
			
		}
%		\vspace{-3mm}
		\caption{University campus: incoming queries and outgoing responses, measured during 3 June to 9 June 2019.}
		\vspace{-5mm}
		\label{fig:InQryUNSW}
	\end{center}
	%\vspace{-2mm}
\end{figure}

\textbf{Traffic Composition of Each Host:}
Now we consider the distribution of normal outbound lookups, unanswered queries, error lookups, and unsolicited inbound responses within each host.
Fig.~\ref{fig:ccdfUnsolicitRespInUNSW} is the stack plot for the top 100 internal IP addresses with the most number of outgoing lookups (more than $35K$ over a week) with no error replies, and each bar represents an individual host.
Seventy-three of them have more than 80\% normal DNS packets in their outbound queries and inbound responses. The major unwanted DNS packet type is error lookups (red shades), such as {\myverb{NameError}}, {\myverb{ServerFailure}} and {\myverb{QueryRefused}}. It might be because of typo error in domain names or malicious DNS activities such as DoS attack or contacting remote attackers using random domain strings \cite{SSchuppenSec2018}.
Unanswered queries (black shades) sent to external IP addresses that do not get a reply back are the second popular reason. Especially for hosts 13, 14, 15, and 19 with $23.5\%$, $23.2\%$, $23.3\%$, and $50.1\%$ such unwanted outgoing queries, respectively -- they are likely to be infected servers or host hackers that generating DNS scans or DoS attacks.
Besides, three university hosts (order 56, 60, and 62) are also suffering from many unsolicited responses, occupying $9.29\%$, $3.96\%$, and $6.92\%$ of their total number of packets for outbound queries and inbound responses. They suffered from small-scale DNS reflection attacks; for example, $99.04\%$ unsolicited responses targeting host 56 are from only one recursive resolver configured by a private company located in China. %101.198.198.198

\subsubsection{Incoming Queries \& Outgoing Responses}\label{sec:InQryOutResp}

Enterprises commonly receive DNS queries from the Internet that are addressed to their authoritative name servers.

It can be seen that two hosts of the University campus (\ie hosts C and D in Fig.~\ref{fig:outgoingresponsesUNSW}) are the dominant contributors to outgoing DNS responses -- we have verified (by reverse lookup) that these hosts are indeed the name servers of the organization.
Interestingly, for both organizations, we observe that a large number of hosts (\ie 197K IP addresses (shown by red shades in Fig.~\ref{fig:incomingqueriesUNSW} for the university network) receive queries from the Internet. Still, a significant majority of them are unanswered (\ie $75.6\%$). These hosts are supposed to neither receive nor respond to incoming DNS queries, highlighting the amount of unwanted DNS traffic that targets enterprise hosts for scanning or DoS purposes.

\begin{figure}[t!]
	%	\vspace{-3mm}
	\begin{center}
		\mbox{
			\hspace{-6mm}
			\subfigure[CCDF: \# Unwanted DNS pkts per host.]{
				{\includegraphics[width=0.25\textwidth]{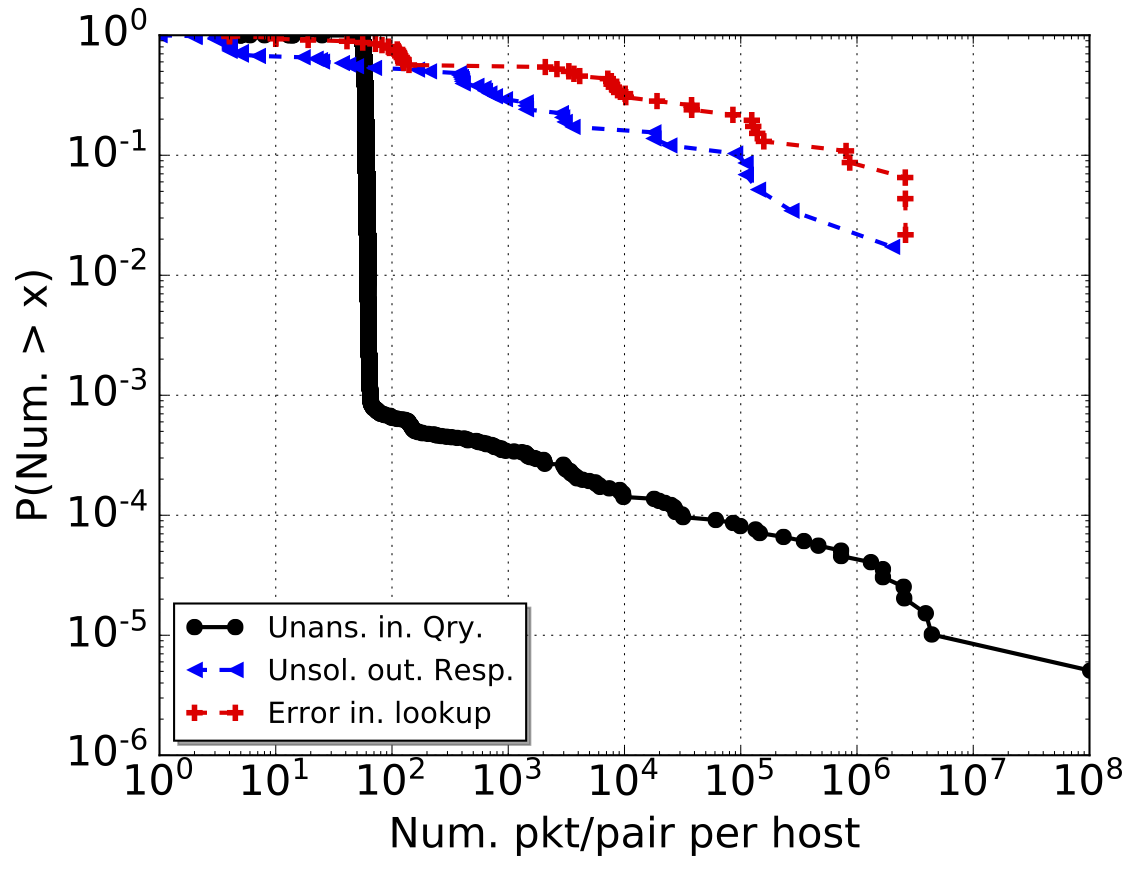}}\quad
				\label{fig:ccdfUansweredQryInUNSW}
			}
			\hspace{-6mm}
			\subfigure[Traffic composition per host.]{
				{\includegraphics[width=0.25\textwidth]{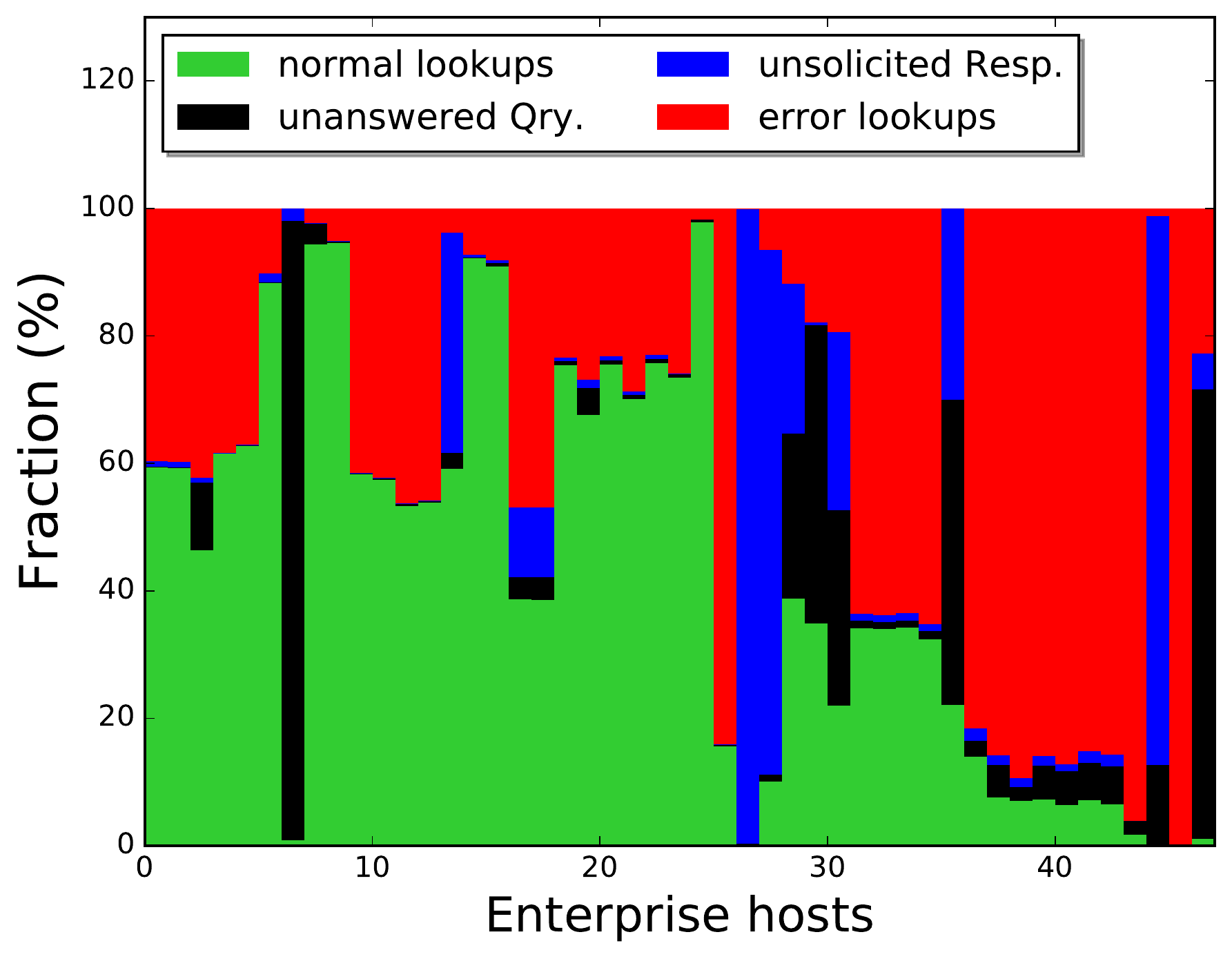}}\quad
				\label{fig:ccdfUnsolicitRespOutUNSW}
			}
		}
		%	\vspace{-3mm}
		\caption{University campus: (a) CCDF of \# unwanted (incoming queries and outgoing responses) DNS packets and (b) their total fractions per host, measured during 3 June to 9 June 2019.}
		%	\vspace{-5mm}
		\label{fig:ccdfUnsolicitRespOutUnansweredQryInUNSW}
	\end{center}
	\vspace{-8mm}
\end{figure}

\textbf{Unwanted DNS Packets Per Host:}
To better understand hosts involved in incoming queries and outgoing responses, we show the distribution of inbound unanswered queries, lookups replied with error responses and unsolicited outgoing responses from hosts inside the two enterprises.

Fig.~\ref{fig:ccdfUansweredQryInUNSW} shows the CCDF plot of the distributions per host for the university campus. 
More than 99\% enterprise IPs (including unassigned IP addresses) received unanswered queries from the Internet. As shown as the black line, almost all IPs are targeted by a small number (\ie less than 100) of such queries over a week -- it indicates active and frequent DNS scans toward the organization. 
Some internal hosts received a massive amount of inbound queries at a high packet rate, located at the tail of the black line in Fig.~\ref{fig:ccdfUansweredQryInUNSW}, are likely to be victims of query flooding attacks. For example, a mixed DNS server (\ie performs as both authoritative name server and local recursive resolver) operated by a school in engineering faculty received $102M$ ($75.5\%$ of all unanswered incoming queries) lookups asking for non-enterprise services such as ``{\myverb{google.com}}'' and ``{\myverb{163.com}}''.

Moreover, 59 hosts sent unsolicited outbound responses (due to server misconfiguration, used as a reflector by internal attackers or packet drop); 47 hosts sent responses with errors (due to typos in domain names by outside users or being as victims in query-based attacks). In Fig.~\ref{fig:ccdfUansweredQryInUNSW}, the hosts that send unsolicited outbound responses are shown as blue dots, and the hosts that send responses with errors are shown as red dots.
The top 3 hosts that sent most of the unsolicited responses ($86.1\%$) are all servers operated by sub-department (verified by reverse lookups), and the organizational IT department does not have knowledge and control over them, highlighting the security blind spots for a large enterprise network.

\textbf{Traffic Composition of Each Host:}
Now we look at the distribution of normal inbound lookups, unanswered queries, outbound error lookups, and unsolicited responses within each host, as shown in Fig.~\ref{fig:ccdfUnsolicitRespOutUNSW} for 47 university hosts that sent outbound responses.
It is clear that only six hosts are associated with more than $80\%$ normal inbound lookup packets, and 45 hosts generated responses with error code other than {\myverb{NoError}}.
Interestingly, $2,083$ out of $2,085$ outbound responses from the 45th host are labeled as error lookups -- it is likely to be an idle authoritative name server, which received irrelevant questions such as ``{\myverb{researchscan541.eecs.umich.edu}}'', ``{\myverb{www.qq.com}}'' and ``{\myverb{www.wikipedia.org}}'' and respond with {\myverb{QueryRefused}}.
Three hosts (ranked 26, 27, and 44 in terms of the number of outgoing responses) are occupied by more than 90\% unsolicited responses. They are all operated by sub-departments and are potential error-configured (such as unsynchronized timing) or reflecting DNS responses for internal attackers, as we observed a significant amount of unsolicited responses for question name {\myverb{miep}} under the deprecated service type {\myverb{ANY}} and other irrelevant to the enterprise zone.
Finally, three internal hosts suffered from a large fraction (more than 50\%) of unanswered queries, especially for the 7th host -- it is the mixed DNS server in engineering faculty as mentioned above, which was consistently under DoS attacks by irrelevant queries. The exhaustion of server resources led to it becoming unresponsive to most incoming queries (and only about 1\% of queries got answered, including relevant and irrelevant questions).

\section{Clustering Enterprise DNS Assets}\label{sec:class}

In this section, we firstly articulate key attributes that can effectively differentiate types of DNS-related enterprise hosts (\S\ref{sec:attributes}). We then develop a unsupervised clustering technique to determine if an enterprise host with a given DNS activity is a ``name server'', ``recursive resolver'', ``mixed DNS server", or a ``regular end-host'' (\S\ref{sec:hostclustering}). We then rank the enterprise DNS servers into ``name server'' and ``recursive resolver'' by their importance, whereas mixed DNS servers are ranked in both types (\S\ref{sec:ServerRanking}). Finally, the regular end-hosts can be further clustered as ``NATed'' or ``not-NATed'' based on their DNS activities as described by the proposed attributes (\S\ref{sec:NAT}).

Our proposed system automatically generates lists of active servers into three categories located inside enterprise networks and rankings in terms of their name server and resolver functionalities, with the real-time DNS data mirrored from the border switch of enterprise networks. 
The system first performs \textit{\textbf{``Data cleansing''}} that aggregates DNS data into one-day granularity and removes unsolicited responses and unanswered queries (\ie step 1); then \textit{\textbf{``Attribute extraction''}} in step 2 computes attributes required by the following algorithms; \textit{\textbf{``Server mapping''}} in step 3 classify DNS assets of various types; and finally \textit{\textbf{``Server ranking''}} in step 4 ranks their criticality. The output is a classification and a ranked order of criticality, which an IT manager can then use to accordingly adjust management and security policies.

\begin{table}[t!]
	\centering
	%	\vspace{-3mm}
	\caption{Samples of host attributes.}
	\label{tab:smapleAttrbt}
	%	\vspace{-2mm}
	\renewcommand{\arraystretch}{1.2}
	\begin{adjustbox}{max width=0.5\textwidth}
		\begin{tabular}{|l|c|c|c|c|}
			%		\begin{tabular}{|l|c|c|c|c|}
			\hline
			\rowcolor[rgb]{ .906,  .902,  .902}			& \textbf{QryFracOut} & \textbf{fracExtSrv} & \textbf{fracExtClient} & \textbf{actvQryOutTime} \\ \hline\hline
			Univ name serv. (host C)        & 0                                & 0                                & 0.26                             & 0                                \\ \hline
			Rsch main name server        & 0                                & 0                                & 0.42                            & 0                                \\ \hline
			Univ rec. resolv. (host A)& 1                                & 0.23                            & 0                                & 1                                \\ \hline
			Rsch main recurs. resolv. & 1                                & 0.43                             & 0                                & 1                                \\ \hline
			Univ mixed DNS Server              & 0.31                             & 0.02                             & 0.03                           & 1                                \\ \hline
			Rsch mixed DNS Server              & 0.23                             & 0.0003                           & 0.0013                          & 1                                \\ \hline
			Univ end-host               & 1                                & 0.00001                          & 0                                & 0.041                            \\ \hline
			Rsch end-host               & 1                                & 0.00001                        & 0                                & 0.25                             \\ \hline
		\end{tabular}
		%	\vspace{-3mm}
	\end{adjustbox}
	%	\vspace{-5mm}
\end{table}

\subsection{Attributes}\label{sec:attributes}
Following the insights obtained from the DNS behavior of various hosts, we now identify attributes that help automatically (a) map a given host to its function including authoritative name server, recursive resolver, mixed DNS server (\ie both name server and recursive resolver), or a regular client; and (b) rank the importance of DNS servers. All attributes are computed from DNS packets' metadata (\ie headers) without inspecting their payload, resulting in a cost-effective inference method.

\subsubsection{Dataset Cleansing} 
We first clean our dataset by removing unwanted (or malicious) records including unsolicited responses and unanswered queries -- it removes the large fraction of unassigned or inactive IP addresses that are only associated with incoming DNS traffic. This is done by correlating the transaction ID of responses with the ID of their corresponding queries. In the cleaned dataset, incoming responses are equal in number to outgoing queries, and similarly for the number of incoming queries and outgoing responses.

\subsubsection{Functionality Mapping} 
As discussed in \S\ref{sec:analysis}, recursive resolvers are very active in terms of queries-out and responses-in, whereas name servers behave the opposite with high volume of queries-in and responses-out. Hence, a host attribute defined by the \textit{query fraction of all outgoing DNS packets (\textbf{QryFracOut})} should distinguish recursive resolvers from name servers. As shown in Table~\ref{tab:smapleAttrbt}, this attribute has a value close to 1 for recursive resolvers and a value close to 0 for name servers.

In addition to recursive resolvers, there are some end-hosts configured to use public resolvers (\eg 8.8.8.8 of Google) that have a non-zero fraction of DNS queries out of the enterprise network. We note that these end-hosts ask a limited number of Internet servers during their activity period whereas the recursive resolvers typically communicate with a larger number of external servers. Thus, we define a second attribute as the \textit{fraction of total number of external servers queried (\textbf{fracExtSrv}) per individual enterprise host}.  As shown in Table~\ref{tab:smapleAttrbt}, the value of this attribute for end-hosts is much smaller than for recursive resolvers.
Similarly for incoming queries, we consider a third attribute as the \textit{fraction of total number of external hosts that initiate query in (\textbf{fracExtClient}) per individual enterprise host}. Indeed, this attribute has a larger value for name servers compared with other hosts, as shown in Table~\ref{tab:smapleAttrbt}.

Lastly, to better distinguish between end-hosts and recursive resolvers (high and low profile servers), we define a fourth attribute as the \textit{fraction of active hours for outgoing queries (\textbf{actvQryOutTime})}. For each host, this attribute indicate the fraction of time it sends outgoing queries. Regular clients have a smaller value of this attribute compared with recursive resolvers and mixed DNS servers, as shown in Table~\ref{tab:smapleAttrbt}.

\subsubsection{Importance Ranking}\label{sec:importanceranking} 
Two different attributes are used to rank the importance of name servers and recursive resolvers respectively. Note that we rank mixed DNS servers within both name servers and recursive resolvers for their mixed DNS behaviour.
For recursive resolvers, we use \textbf{\textit{QryFracHost}} defined as the \textit{fraction of outgoing queries} sent by each host over the cleaned dataset. And for name servers, we use \textit{\textbf{RespFracHost} as the fraction of outgoing responses} sent by each host. 

\begin{table}[t!]
	\centering
	%	\vspace{-3mm}
	\caption{University campus: host clusters (3 June 2019).}
	%\vspace{-2mm}
	\label{tab:clusterUNSW}
	\renewcommand{\arraystretch}{1.2}
	\begin{adjustbox}{max width=0.5\textwidth}	
		\begin{tabular}{|l|c|c|c|c|c|}
			\hline
			\rowcolor[rgb]{ .906,  .902,  .902}		& \textbf{Count} & \textbf{QryFracOut}& \textbf{fracExtSrv} & \textbf{fracExtClient}&\textbf{actvQryOutTime} \\ \hline\hline
			name server        & 24                                                                             & 0.0004                           & 1e-5                     & 0.03                             & 0.04                             \\\hline
			recursive resolver & 21                                                                             & 0.99                             & 0.04                             & 6e-5                             & 0.77                             \\\hline
			mixed DNS srv.        & 22                                                                             & 0.57                             & 0.008                             & 0.01                             & 0.64                             \\\hline
			end-host      & 2,518                                                                           & 1.00                                & 3e-5                          & 0.00                                & 0.24                             \\ \hline
		\end{tabular}
	\end{adjustbox}
	\vspace{-2mm}
\end{table}

\subsection{Host Clustering}\label{sec:hostclustering}
We choose unsupervised clustering algorithms to perform the grouping and classification process because they are a better fit for datasets without ground truth labels but nevertheless exhibit a clear pattern for different groups/clusters.

\subsubsection{Selecting Algorithms}
We considered 3 common clustering algorithms, namely Hierarchical Clustering (HC), K-means and Expectation-maximization (EM). 
%HC
HC is more suitable for datasets with a large set of attributes and instances that have logical hierarchy (\eg genomic data). In our case however, hosts of enterprise networks do not have a logical hierarchy and the number of attributes are relatively small, therefore HC is not appropriate.
%K-means
K-means clustering algorithms are distance-based unsupervised machine learning techniques. By measuring the distance of attributes from each instance and their centroids, it groups data-points into a given number of clusters by iterations of moving centroids. 
In our case there is a significant distance variation of attributes for hosts within each cluster (\eg highly active name servers or recursive resolvers versus low active ones) which may lead to mis-clustering.

%EM is the best for our case
The EM algorithm is a suitable fit in our case since it uses the probability of an instance belonging to a cluster regardless of its absolute distance. It establishes initial centroids using a K-means algorithm, starts with an initial probability distribution following a Gaussian model and iterates to achieve convergence. This mechanism, without using absolute distance during iteration, decreases the chance of biased results due to extreme outliers. Hence, we choose an EM clustering algorithm for \textit{``DNS Host Clustering Machine''}.

\subsubsection{Number of Clusters}
\begin{figure}[t!]
%	\vspace{-3mm}
	\begin{center}
		{\includegraphics[width=0.43\textwidth,height=0.3\textwidth]{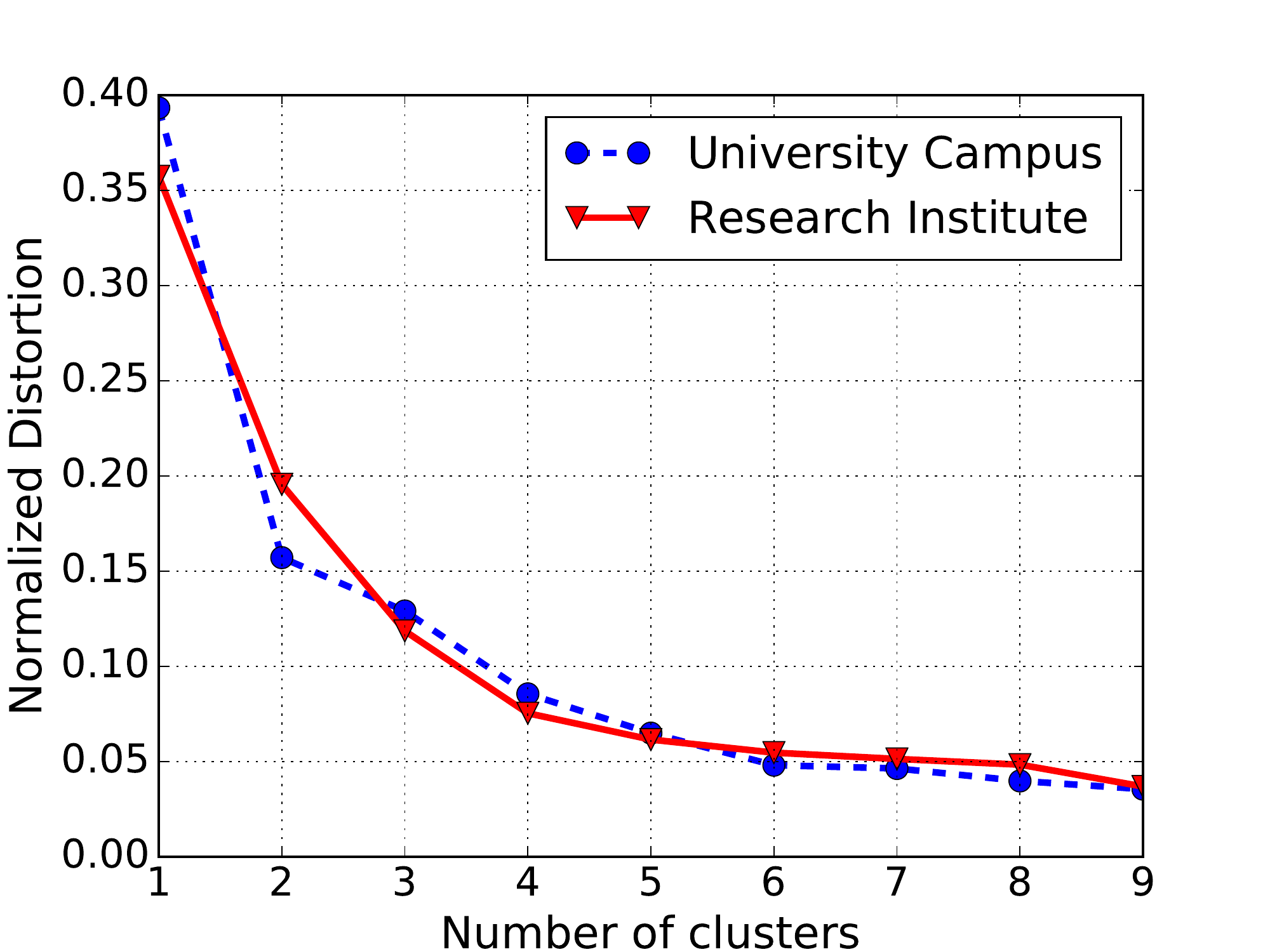}}
		%		\vspace{-1mm}
		\caption{Elbow method: evaluating number of clusters.}
		%	\vspace{-1mm}
		\label{fig:numOfClusters}
	\end{center}
		\vspace{-6mm}
\end{figure}

Choosing the appropriate number of clusters is the key step in clustering algorithms.
As discussed earlier, we have chosen four clusters based on our observation of various types of servers. 
One way to validate the number of clusters is with the ``elbow'' method. The idea of the elbow method is to run k-means clustering on the dataset for a range of k values (say, k from 1 to 9 as shown in Fig.~\ref{fig:numOfClusters}) that calculates the sum of squared errors (SSE) for each value of k. The error decreases as k increases; this is because as the number of clusters increases, the SSE becomes smaller so the distortion also gets smaller. 
The goal of the elbow method is to choose an optimal k around which the SSE decreases abruptly (\ie ranging from $3$ to $5$ in our results, hence, $k = 4$ clusters seems a reasonable value for both the university and the research institute).

\subsubsection{Clustering Results}

\begin{table}[t!]
	\centering
	%	\vspace{-3mm}
	\caption{Research institute: host clusters (3 June 2019).}
	\label{tab:clusterCSIRO}
	%	\vspace{-2mm}
	\renewcommand{\arraystretch}{1.2}
	\begin{adjustbox}{max width=0.5\textwidth}	
		\begin{tabular}{|l|c|c|c|c|c|}
			\hline
			\rowcolor[rgb]{ .906,  .902,  .902}			& \textbf{Count} & \textbf{QryFracOut} & \textbf{fracExtSrv} & \textbf{fracExtClient} & \textbf{actvQryOutTime} \\ \hline\hline
			name server        & 13                                                                             & 0.00                       & 0.00                         & 0.07                             & 0.00                             \\\hline
			recurs. resolv. & 25                                                                             & 1.00                             & 0.03                             & 0.00                          & 0.86                                \\\hline
			mixed DNS srv.     & 2                                                                              & 0.81                             & 0.05                            & 0.04                            & 0.54                            \\\hline
			end-host      & 245                                                                            & 1.00                               &5e-4                           & 0.00                                & 0.17                             \\ \hline
		\end{tabular}
	\end{adjustbox}
		\vspace{-2mm}
\end{table}

\begin{figure*}[t!]
	\hspace{-5mm}
	%	\vspace{-8mm}
	\begin{center}
		%		\vspace{-5mm}
		\mbox{
			%	\vspace{-5mm}
			\hspace{-3mm}
			\subfigure[Univesity campus.]{
				%	\hspace{-5mm}
				{\includegraphics[width=0.55\textwidth]{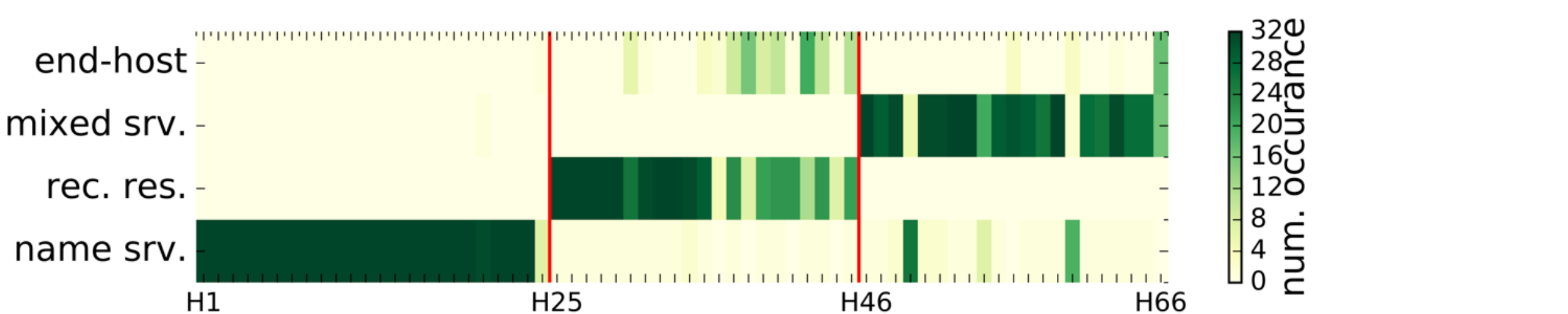}}
				\label{fig:clusteringWkUNSW}
				%		\vspace{-10mm}
			}
			%		}
			\hspace{-13mm}
			%		\mbox{
			%	\hspace{-5mm}
			%		\vspace{-8mm}
			\subfigure[Research institute.]{
				\hspace{-8mm}
				{\includegraphics[width=0.65\textwidth]{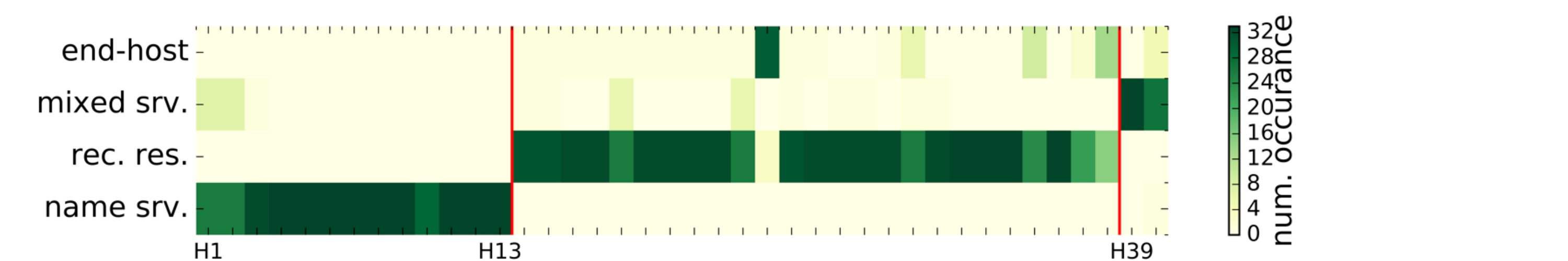}}
				\label{fig:clusteringWkCSIRO}
			}
		}
		%		\vspace{-3mm}
		\caption{Hosts clustering results across 32 days.}
		\vspace{-4mm}
		\label{fig:clusteringWk}
	\end{center}
		\vspace{-4mm}
\end{figure*}

We tuned the number of iterations and type of covariance for our clustering machine to maximize the performance in both enterprises.
Tables~\ref{tab:clusterUNSW} and \ref{tab:clusterCSIRO} show the number of hosts identified in each cluster based on data from 3 June 2019. We also see the average value of various attributes within each cluster.
%name servers
For the cluster of name servers, \textit{QryFracOut} approaches 0 in both organizations (some name servers performed outbound DNS lookups for its own operational purposes), highlighting the fact that almost all outgoing DNS packets from these hosts are responses rather than queries, which matches with the expected behavior. Having a high number of external clients served also indicates the activity of these hosts -- in the University campus and research institute respectively 24 and 13 name servers collectively serve $81.6$\% and $91$\% (\ie $24\times3.4\%$ and $13\times7\%$) of external hosts.

%recursive resolvers
Considering recursive resolvers in Tables~\ref{tab:clusterUNSW} and \ref{tab:clusterCSIRO}, the average \textit{QryFracOut} is close to 1 for both organizations as expected. It is seen that some of these hosts also answer incoming queries (from external hosts) possibly due to their mis-configuration. However, the number of external clients served by these hosts is very small (\ie less than 5 per recursive resolver) leading to an average fraction near 0. Also, looking at the number of external servers queried (\ie \textit{fracExtSrv}), the average value of this attribute for recursive resolvers is reasonably high, \ie 21 and 25 hosts in the University and the research network respectively contribute to 83\% and 89\% of total \textit{fracExtSrv} -- this is also expected since they commonly communicate with public resolvers or authoritative name servers on the Internet.

%MX and End host
Hosts clustered as mixed DNS servers in both organizations have a moderate value of the \textit{QryFracOut} attribute (\ie $0.57$ and $0.81$ for the University and the research network respectively) depending on their varying level of inbound/outbound DNS activity. Also, in terms of external clients and servers communicated with, the mixed servers lie between name servers and recursive resolvers. Lastly, regular end-hosts generate only outbound DNS queries (\ie \textit{QryFracOut} equals to $1$), contact a small number of external resolvers, and are active for shorter duration of time over a day (\ie \textit{actvQryOutTime} less than $0.5$).

\subsubsection{Interpreting the Confidence of Clustering}
Our clustering algorithm also generates a confidence level as an output. This can be used as a measure of reliability for our classifier. If adequate information is not provided by attributes of an instance then the algorithm will decide its cluster with a low confidence level -- this can be interpreted as an ``unknown'' cluster.
The average confidence level of the result clustering is $98.13$\% for both organizations, with more than $99$\% of instances classified with a confidence-level of more than $85$\%. This indicates the strength of our host-level attributes, enabling the algorithm to cluster them with a very high confidence-level.

\subsubsection{Server Clusters Across 32 Days}

We now check the performance of our clustering algorithm over 32 days. Fig.~\ref{fig:clusteringWk} shows a heat map for clusters of servers. Columns %(or x-axis) 
list server hosts that were identified in Tables~\ref{tab:clusterUNSW} and \ref{tab:clusterCSIRO} (\ie 66 hosts in the University network and 40 hosts in the research network). Rows display the cluster into which each server is classified. The color of each cell depicts the number of days (over 32 days) that each host is identified as the corresponding cluster -- dark cells depict a high number of occurrences (approaching 32), while bright cells represent a low occurrence closer to 0. 

In the University network we identified 25 name servers, shown by H1 to H25 in Fig.~\ref{fig:clusteringWkUNSW}; the majority of which are repeatedly classified as a name server over 32 days, thus represented by dark cells at their intersections with the bottom row, highlighting the strong signature of their profile as a name server. Exceptions is H25, which was only active for 7 days as name server and 1 day as end-host. It is an IP address belong to school of physics under department of science, as verified by reverse lookups.

Among 21 recursive resolvers of the university campus, shown by H25 to H46 in Fig.~\ref{fig:clusteringWkUNSW}; 7 of them (including hosts A and B in Fig.~\ref{fig:outQryUNSW}) are consistently classified as recursive resolver, and the rest are re-classified as end hosts (due to their varying activity). Lastly, 20 mixed servers, shown by H46 to H66 in Fig.~\ref{fig:clusteringWkUNSW}, are classified consistently though their behavior sometimes is closer to a end-host or a name server. 

Our results from the Research Institute network are fairly similar -- Fig.~\ref{fig:clusteringWkCSIRO} shows that hosts H1-H13 are consistently classified as name servers, while hosts H14-H38 are recursive resolvers and H39-H40 are mixes servers. 
Unlike the University Campus, 9 recursive resolvers are classified as mixed-server from 1 to 6 days. They are owned by business units in the organization, revealing the dynamicity of their DNS infrastructures.

\subsubsection{IT Verification of Clustering Results}
The IT department in both organizations verified the top-ranked DNS resolvers and name-servers found across the 32 days, meaning 100\% accuracy for ground-truth DNS assets, as they are directly configured and controlled by the IT departments.
Additionally, we revealed unknown servers configured by departments of the two enterprises (we verified their functionality by reverse DNS lookup and their IP range allocated by IT departments).
Interestingly, 3 of the name-servers our method identified were involved as reflectors in a DNS amplification attack, and IT was able to confirm that these were managed by affiliated entities (such as retail stores that lease space and Internet connectivity from the University) - this clearly points to the use of our system in identifying and classifying assets whose security posture the network operators themselves may not have direct control over.

\subsection{Server Ranking}\label{sec:ServerRanking}

Our system discovered 46 authoritative name servers and 43 recursive resolvers in the University (a mixed DNS server are treated as both name server and recursive resolver), and 15 authoritative name server and 27 recursive resolvers at the Research Institute. However, only 6 top ranked DNS servers, in each organization, contribute to more than 90\% of outgoing queries and responses. Servers ranking provides network operators with the popularity of their DNS assets.

\begin{figure}[t!]
	\begin{center}
		\mbox{
			\hspace{-7mm}
			\subfigure[University campus.]{
				{\includegraphics[width=0.25\textwidth]{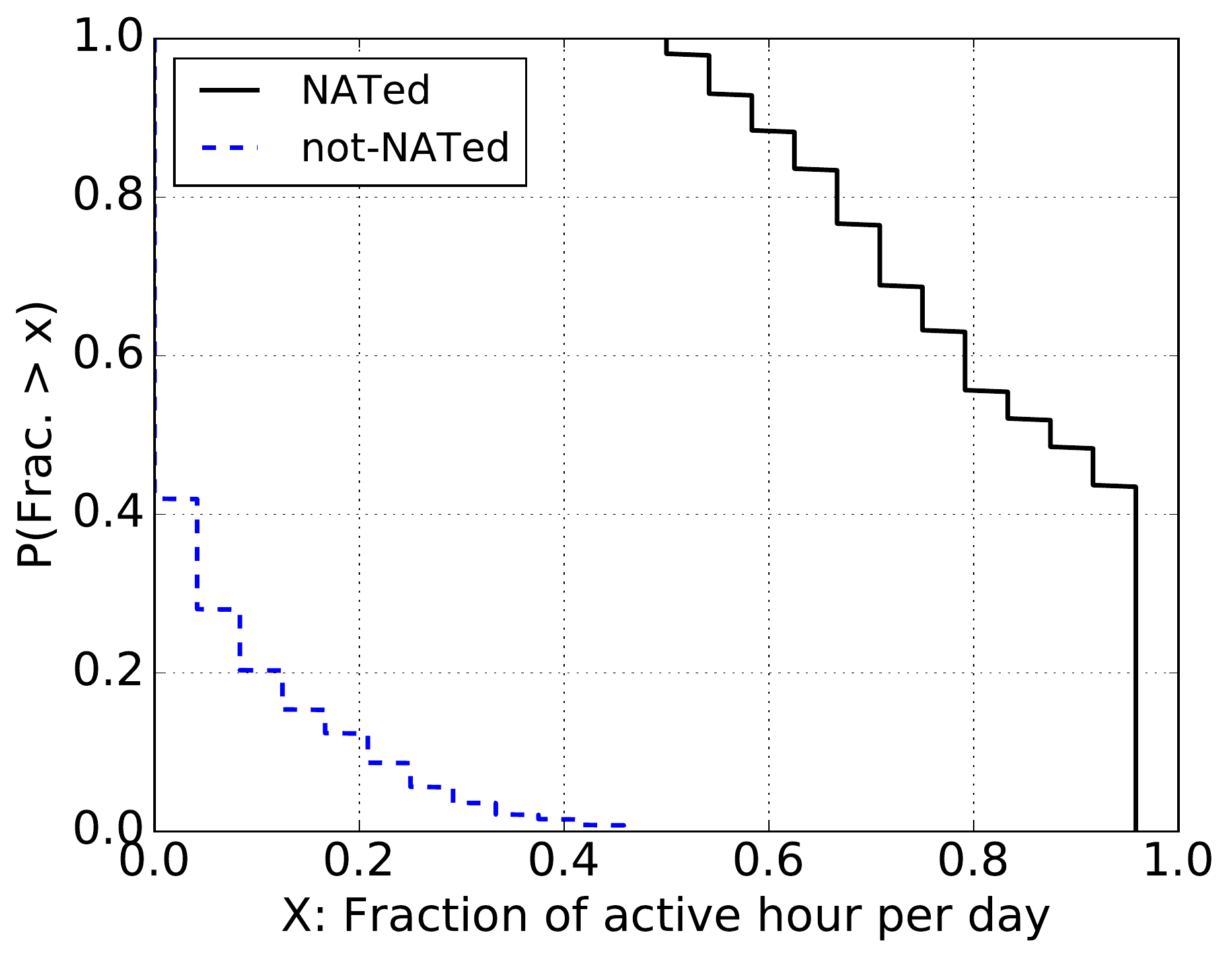}}\quad
				\label{fig:endhostFracUniv}
			}
			\hspace{-7mm}
			\subfigure[Research institute.]{
				{\includegraphics[width=0.25\textwidth]{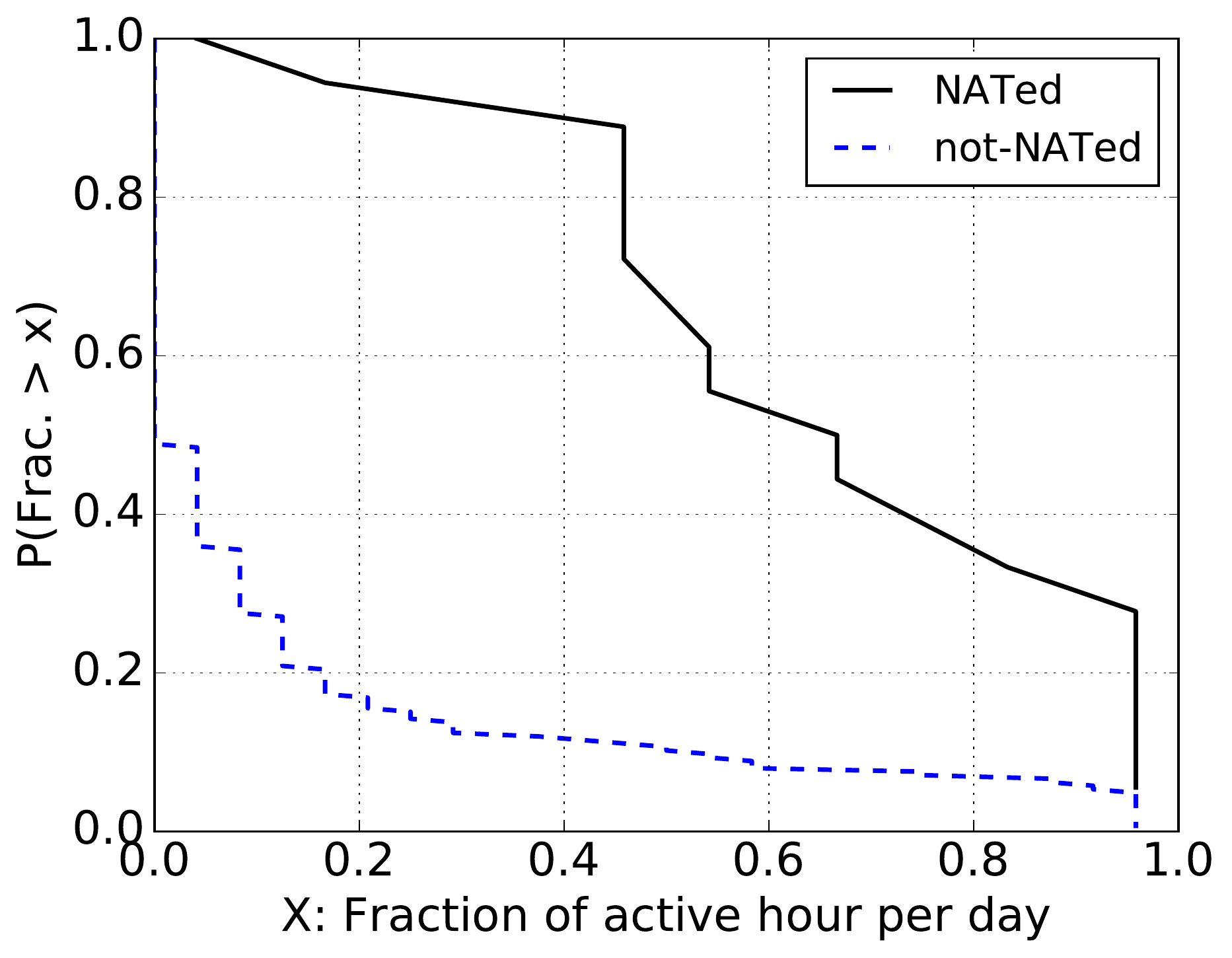}}\quad
				\label{fig:endhostFracRsch}
			}
		}
		\vspace{-2mm}
		\caption{CCDF: fraction of active hour per day for NATed and not-NATed end-host IP addresses.}
		\vspace{-3mm}
		\label{fig:endhostFrac}
	\end{center}
	\vspace{-3mm}
\end{figure}

\subsection{Clustering of End-hosts: NATed or Not?}\label{sec:NAT}

To draw more insights we further applied our clustering algorithm (using the same attributes introduced in \S\ref{sec:attributes}) to IP address of end-hosts, determining whether they are behind a NAT gateway or not (\ie two clusters: NATed and not-NATed). In both networks, all WiFi clients are behind NAT gateways. Additionally, some specific departments of the two enterprises use NAT for their wired clients too. 
We verified our end-host clustering by reverse lookup for each enterprise network. Each NATed IP address has a corresponding domain name  in specific forms configured by IT departments. For example the University campus wireless NAT gateways are associated with domain-names as 
``{\myverb{SSID-pat-pool-a-b-c-d.gw.unsw.edu.au}}'', where ``{\myverb{a.b.c.d}}'' is the public IP address of the NAT gateway, and ``{\myverb{SSID}} is the the WiFi SSID for the University campus network.
Similarly, in the Research institute NAT gateways use names in form of ``{\myverb{c-d.pool.rsch-primary-domain}}'' where ``{\myverb{c.d}}'' is the last two octets of the public IP address of the NAT gateway in the Research institute.

%result
On 3rd June 2019, our end-host clustering shows that 337 and 42 of end-hosts IP addresses are NATed in the University campus and the Research institute respectively.
%accuracy evaluation
We note that the two clusters of end-hosts are distinguished primarily by two attributes, namely \textit{actvTimeFrac} -- a NATed IP address (representing a group of end-hosts) is expected to have a longer duration of DNS activity compared to a not-NATed IP address (representing a single end-host), as illustrated in Fig.~\ref{fig:endhostFrac}, and \textit{numExtSrv} -- a NATed IP address is expected to have more than one queried public DNS resolvers, as it is represent many individual hosts each connect with their selected resolvers on the Internet. All classified not-NATed hosts contacted less than 10 external DNS servers in both organizations during 3rd June, while $54\%$ and $26\%$ NATed IPs in the university and research institute were queried more than 10 public servers.

%verification by Domain names
We verified their corresponding domain names configured by their IT departments.
Some IPs with domain-names of NAT gateways are incorrectly classified as not-NATed end-hosts. This is because their daily DNS activity was fairly low, \ie less than an hour with only one external resolver contacted. On the other hand, not-NATed end-hosts with long duration of DNS activity (\ie almost the whole day) were misclassified.   
Verifying end-hosts classified as NATed, $77.2\%$ of them in the University campus and and $75.0\%$ in the Research institute have corresponding domain-names as for NAT gateways allocated by IT departments. For end-hosts classified as not-NATed, $91.1\%$ and $93.6\%$ in the respective two organizations do not map to any organizational domain-names.

\begin{figure}[t!]
	\begin{center}
		\mbox{
			\hspace{-7mm}
			\subfigure[University campus.]{
				{\includegraphics[width=0.25\textwidth]{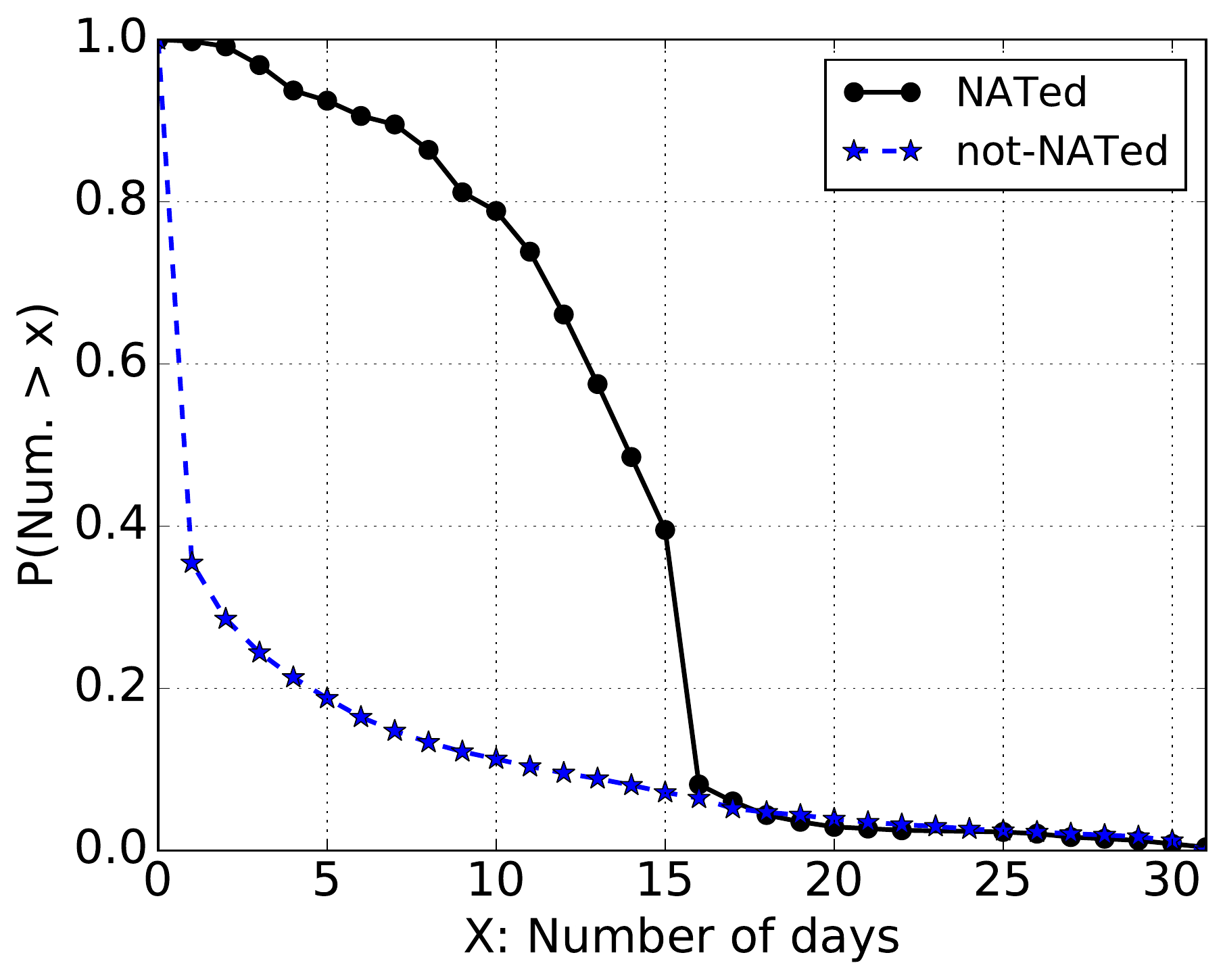}}\quad
				\label{fig:WeekConsistencyUniv}
			}
			\hspace{-7mm}
			\subfigure[Research institute.]{
				{\includegraphics[width=0.25\textwidth]{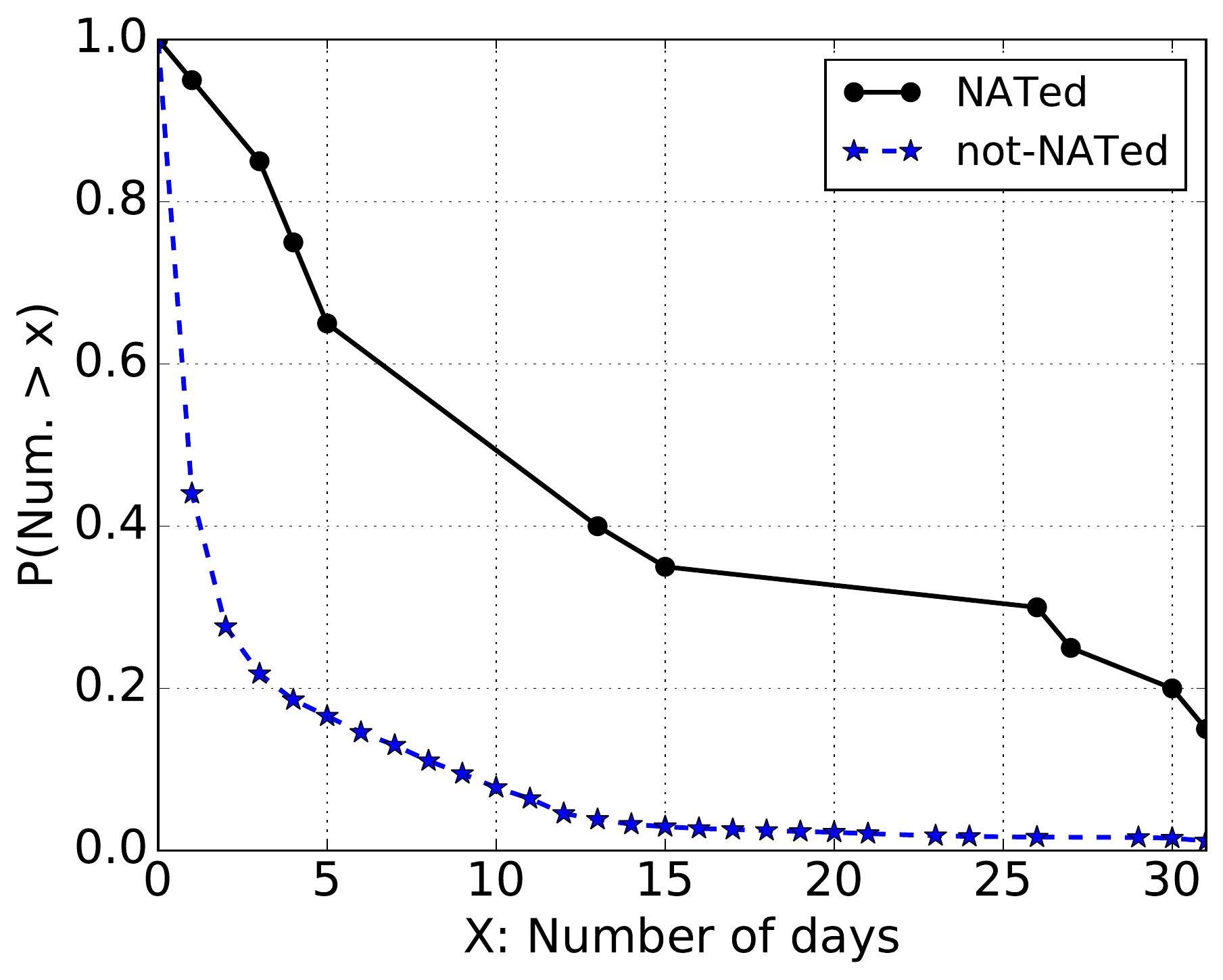}}\quad
				\label{fig:WeekConsistencyRsch}
			}
		}
		\vspace{-1mm}
		\caption{CCDF: Consistency of end-hosts clustering across 32 days.}
		%	\vspace{-3mm}
		\label{fig:WeekConsistency}
	\end{center}
	\vspace{-8mm}
\end{figure}

Looking into the consistency of end-hosts clustering across 32 days, we note that more than $90\%$ end-hosts in the University campus are consistently labeled as NATed over 7 days (as show in Fig.~\ref{fig:WeekConsistencyUniv}). $52\%$ end-hosts are classified as NATed from 7 days to 15 days. Those IP addresses are owned by sub-departments in the university, and re-shuffled within their subnets by the organizational DHCP servers periodically. 
As for the University IP addresses get classified as not-NATed (\eg desktops with public IP addresses through wired connection), majority ($63\%$) of them only appear once during 32 days. It is because of their low-profile activities and daily IP re-shuffling.

Similar observations were obtained from the research institute (shown in Fig.~\ref{fig:WeekConsistencyRsch}) , except there are 5 IP addresses appeared as NATed across the 32 days -- they belongs to IT infrastructures controlled by critical scientific basements such as Australia Telescope National Facilities, which are separated controlled with more freedom thus not affected by periodically DHCP reallocation.

\section{Monitoring DNS Asset Health}\label{sec:health}
Having shown how DNS assets in an enterprise network can be identified and classified based on their network behavior, we now extend the study to monitor their health continuously. The objective is to detect \textit{anomalous} behavior, indicating that the asset is being misused or attacked, and identify the root cause of such deviations in behavior. 
We begin in \S\ref{sec:motivatingExample} by providing two examples of observable anomalies from our dataset -- one attributable to poor configuration and the other subject to a DDoS attack. Inspired by these examples, in \S\ref{sec:metric} we develop a set of \textit{health metrics} that can track the behavior of each asset along various dimensions, and in \S\ref{sec:typeanomalies} develop a method to label and warn anomalous behaviors based on these health tracking metrics. Finally, in \S\ref{sec:insight} we apply our methods to the 32-day dataset from the two organizations and present results into misuse and attack patterns detected by our methods.

\subsection{Examples Illustrating Anomalous DNS Asset Behavior}\label{sec:motivatingExample}
By manually inspecting our dataset, we could identify several behavioral patterns that seemed unusual. Therefore, we begin by providing a couple of illustrative examples of anomalous behavior and subsequently develop methods to automatically detect misbehaviors by tracking various health metrics.

\textbf{Example 1 -- DNS Misuse:} We found that one of the Authoritative DNS servers at the Research Institute was responding with an unexpected high number of ``{\myverb{NXDOMAIN}}'' messages, indicating that corresponding queried domain names do not exist. Manual  investigation revealed that those queried names were irrelevant (\eg ``{\myverb{www.taobao.com}}'') to the organization. In fact, almost a third of DNS queries to this server were irrelevant; still, it was responding. We also found that about $15$\% of incoming queries were malformed (\eg ``{\myverb{com}}''), to which the server responded with an ``{\myverb{NXDOMAIN}}'' message. This example demonstrates how a poorly configured server can behave outside its intended function. Such a vulnerability exposes the server to attackers who aim to launch a denial-of-service attack or use it as a reflector for attacking others.

\textbf{Example 2 -- DNS Flood Attack:} We found one of the Authoritative DNS Servers in the University dataset to show a sustained $142\%$ increase in inbound query rates over a 10-day period (7-Jun 0:17AM till 17-Jun 4:43PM). Upon closer inspection, we noted 3.3M queries, all with the same query name ``{\myverb{aids.gov}}'', had come during this period from 974 external sources -- typically each external source launched around 300 queries within a 20 second period, and then went idle. The DNS server was unable to keep up with the high rate of requests, and was able to service only about $70\%$ of incoming queries. Further, $40.9\%$ of the responses during this period were irrelevant to the organization, while $21.6\%$ were with the ``{\myverb{NXDOMAIN}}'' error.

\subsection{DNS Traffic Health Metrics}\label{sec:metric}
Having seen some examples of poor behavior from DNS servers, we now propose several metrics that can be used to track the health of each DNS asset in the organization. We categorize them into service, functional, network, and volumetric behaviors.

\textbf{Service Behavior:}
From a border perspective, authoritative name servers are expected to only serve DNS queries seeking to resolve domains relevant to the enterprise. Conversely, recursive resolvers should only send outbound queries for domains outside of the enterprise -- queries for internal domains are internally sent to the enterprise authoritative name servers without crossing the network border. We therefore define {\myverb{Non-Enterprise Lookup Fraction (NELF)}} as the fraction of query names that are irrelevant to the enterprise services. A properly configured authoritative name server should have {\myverb{NELF}} of 0, while for a recursive resolver this metric should be 1.

\textbf{Functional Behavior:}
Under ideal conditions, responses of a properly functioning DNS server are expected to carry ``{\myverb{NoError}}'' as response code. However, a DNS query may fail due to some errors, such as the domain name queried does not exist, an answer cannot be given, or  the server refuses to answer due to policy.
Therefore, we define {\myverb{Lookup Error Fraction (LEF)}} for a DNS server as the fraction of its responses that carry an error code -- a value significantly larger than 0 indicates potential misbehavior.

\textbf{Network Behavior:}
Under normal circumstances a query is associated with a response. However, the network trace often reveals inbound responses with no outbound queries (\eg a reflective attack to a victim whose IP address was spoofed), as well as outbound queries with no inbound response (\eg a malicious internal host launching a DoS attack via the DNS cache/proxy). To track such anomalous network behavior, we define the Query Service Ratio Inbound (QSRI), \ie ratio of outbound responses to inbound queries, and Query Service Ratio Outbound (QSRO), \ie ratio of inbound responses and outbound queries. All DNS assets should ideally have these two metrics as 1, showing the balanced profile of queries and responses.

\textbf{Volumetric Behavior:}
Sudden increases in the rates of DNS packets can indicate that attacks are targeting enterprise assets. We, therefore, track the inbound and outbound rates of queries and responses for each DNS asset over each epoch (of one hour), respectively {\myverb{QryRateIn (QRI)}}, {\myverb{RespRateOut (RRO)}}, {\myverb{QryRateOut (QRO)}}, and {\myverb{RespRateIn (RRI)}}, and flag those epochs in which the rate shows an increase above a prescribed threshold value (discussed below), which could be indicative of volumetric attacks.

\begin{figure}[t!]
%	\vspace{-1mm}
	\begin{center}
		{\includegraphics[width=0.5\textwidth]{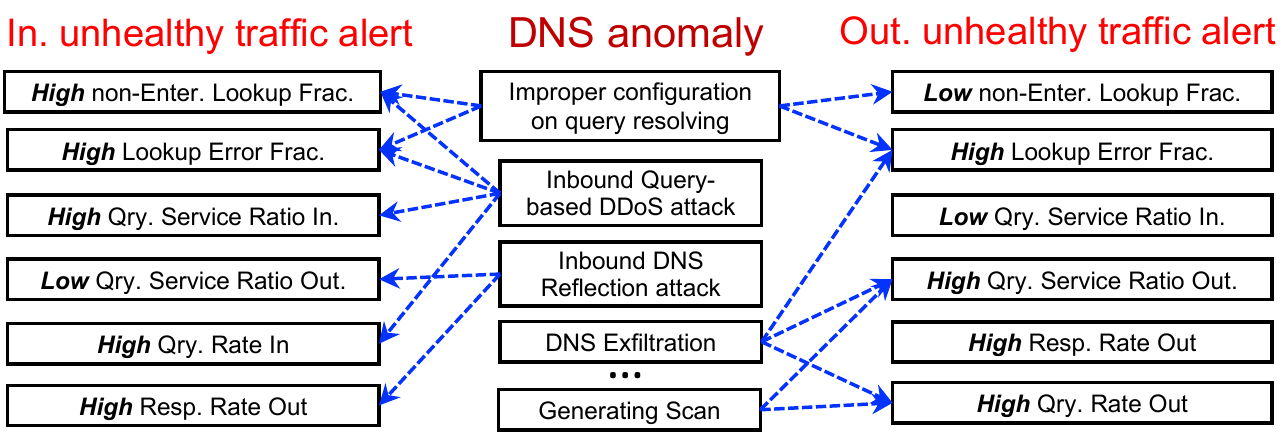}}
	%	\vspace{-2mm}
		\caption{Examples of observed DNS anomalies and their corresponding health alerts.}
		%		\vspace{-4mm}
		\label{fig:MappingAnomalyTypeTrafficHealth}
	\end{center}
	\vspace{-4mm}
\end{figure}

\subsection{Using Health Metrics to Detect Anomalies}\label{sec:typeanomalies}
Using the health metrics identified above, we build a simple mechanism to detect and alert various anomalous behaviors of DNS assets. The set of anomalies we consider in this paper, illustrated pictorially in Fig.~\ref{fig:MappingAnomalyTypeTrafficHealth}, include:
\begin{itemize}
	\item {\textbf{Misconfiguration:}} Consider an authoritative DNS server that has been poorly configured and resolves queries for domains that it has no authority over (\ie do not belong to the enterprise). The exploitation of this by attackers (\eg as a reflector) will manifest in an alert when the NELF metric becomes high, while LEF could also be high (in case the queries are malformed or non-existent). Conversely, a misconfiguration alert is triggered when the NELF metric falls below a threshold value for a poorly configured recursive DNS resolver.
	\item {\textbf{DDoS Attack:}} A distributed denial-of-service attack on an enterprise DNS server will manifest in the form of a volumetric rise in QRI, potentially accompanied by a high value in NELF and/or LEF. 
	Most queries in DDoS attacks tend to be either fixed or random domains instead of customizing query names specific to the victim enterprise.
	\item {\textbf{DNS Exfiltration:}} An infected enterprise host, attempting to exfiltrate data via DNS, will cause QRO to rise, potentially accompanied by unanswered queries (rise in QSRO) and/or lookup errors (rise in LEF). A combination of these metrics can be used as triggers to conduct a deeper investigation into exfiltration, \eg using the method developed by \cite{JAhmedTNSM2020}. One may argue that QRO is expected to be relatively high for legitimate recursive resolvers. Therefore, we infer from a combination of metrics, each with specific thresholds (value ranges) to cater for some reasonable deviations (discussed in \S\ref{sec:insight}).
	\item {\textbf{Scans:}} The presence of malware in the enterprise that performs outbound scans can be detected by monitoring for a rise in outbound queries (QRO), potentially accompanied by unanswered queries (rise in QSRO) and/or lookup errors (rise in LEF). 
\end{itemize}

In what follows we continuously track the health metrics of the various DNS assets identified in the two enterprise networks by our earlier clustering algorithm, and evaluate our ability to identify anomalous behaviors indicative of misconfigurations and/or attacks.
Note that our proposed metrics and alerts from DNS behavioral monitoring could be consumed by SIEM platforms and/or combined with security appliances to verify whether an enterprise host is indeed involved in malicious communications or not. Such combined inferences are beyond the scope of this paper.

\begin{table}[t!]
%	\vspace{-3mm}
	\caption{Alerts and occurrence frequency (in the fraction of epochs) for our two example DNS assets.}
	\centering
	\label{tab:exampleAlert}
%	\vspace{-2mm}
\begin{adjustbox}{max width=0.48\textwidth}
	\renewcommand{\arraystretch}{1.2}
	\begin{tabular}{|c|c|c|cc|}
		\hline
\rowcolor[rgb]{ .906,  .902,  .902}	\textbf{Direction}& \textbf{Profile}&	\textbf{Alert}   &  \textbf{Example 1}& \textbf{Example 2}  \\ \hline\hline
		Inbound $\downarrow$&Service&high NELF   &\cellcolor{red!25}  83.7\% & \cellcolor{red!25} 85.6\%  \\ \hline
		$\downarrow$&Functional&	high LEF     & 6.6\% &0.1\%    \\ \hline
		$\downarrow$&Network&	low QSRI     &  \cellcolor{red!25} 94.5\% & \cellcolor{red!25} 15.6\%  \\ \hline
		$\downarrow$&Network&	high QSRO    & 0.1\% & 0.3\% \\ \hline
		$\downarrow$&Volmetric&	high QRI& 0.0\% & \cellcolor{red!25} 29.7\% \\ \hline
		$\downarrow$&Volmetric&	high RRI & 5.1\%  & 9.8\%  \\ \hline\hline
		
		Outbound $\uparrow$&Service&	low NELF & 0.0\% & 0.0\% \\ \hline	
		$\uparrow$ &Functional&high LEF & \cellcolor{red!25} 100.0\% & \cellcolor{red!25} 20.2\% \\ \hline
		$\uparrow$ &Network&	high QSRI      &  0.8\% &0.3\% \\ \hline
		$\uparrow$ &Network&	low QSRO    &  0.0\%  &\cellcolor{red!25} 84.4\%  \\ \hline	
	    $\uparrow$ &Volumetric&	high RRO  & 5.6\% & 7.7\% \\ \hline
		$\uparrow$ &Volumetric&	high QRO & 1.0\% & \cellcolor{red!25} 31.1\%\\ \hline	
	\end{tabular}
\end{adjustbox}
	\vspace{-5mm}
\end{table}

\subsection{Insights in two Enterprise Networks}\label{sec:insight}
We applied the proposed traffic health metrics to our 32-day DNS traces captured from both organizations, comprising the assets as identified earlier in Tables~\ref{tab:clusterUNSW} and \ref{tab:clusterCSIRO} for the University (67 DNS assets) and Research Institute (40 DNS assets) respectively. The metrics are computed each epoch (of one hour), and our first step is to identify epochs wherein the health metrics deviate significantly from their expected values. 
In general, DNS assets in the University raise more alerts than the research institute. In order to limit the number of alerts, we choose a margin value that is at the elbow points in a curve, which is at around the $30$\% mark. This is also consistent with the threshold values used by many state-of-the-arts security appliances, \eg from Palo Alto \cite{PaloAltoBestPractice2018}, Fortinet \cite{FortinetFirewallDDoS2018} and Cisco \cite{CISCOFirewallDDoS2018}. While organizations are free to tune the threshold alerting values for each health metric to suit their environments, in this work for simplicity we will maintain it at $30$\%. In what follows we first examine two DNS assets that exhibited high rates of alerts (as shown in Table~\ref{tab:exampleAlert}), followed by a general overview of alerts across the two organizations. We then design an inference engine that combines the health metric alerts and deduces the nature of the underlying anomaly causing these alerts using the relationships identified earlier in \S\ref{sec:typeanomalies}. 

\textbf{Example 1:}
A DNS server in the University Law Department serves as both authoritative name server and recursive resolver. It exhibited unhealthy elevated NELF metric for $83.7$\% of epochs, and unhealthy depressed QSRI for $94.5$\% of epochs, indicating its \textbf{misconfiguration} was being exploited by attackers for a potential \textbf{DDoS attack}. Queries for ``{\myverb{d.c.b.a.in-addr.arpa}}'' were coming from many external IP addresses, and the server was responding to a vast majority (over 90\%) of them, thereby wasting its resources. The asset also exhibited many epochs ($6.6$\%) of unhealthy LEF metric, indicating that it was proxying \textbf{scans}. On 29-Jun, this server sent queries to 131 external IP addresses, of which 18 responded -- this asset is likely being utilized as a proxy to perform slow reconnaissance scans to discover availability of DNS servers on the Internet, as explored in detail in our other work \cite{MLyuTNSM2021}.

\textbf{Example 2:}
A DNS server in the University Engineering Department also exhibited many inbound health alerts, such as high NELP for $85.6$\% of epochs, low QSRI for $15.6$\% of epochs, and high QRI for $29.7$\% of epochs. Investigation confirmed that it was \textbf{misconfigured} and exploited by attackers using it to launch reflection attacks with queries for domain names such as {\myverb{dnsscan.shadowserver.org}}, {\myverb{researchscan541.eecs.umich.edu}}\footnote{Although some of the domain names seem to be designed for research-based scans, they are indeed misused by malicious actors to launch the relection attacks on 30-Jun as discussed in Example 2.}, and {\myverb{nil}}. The server was also giving outbound health alerts for high LEF, QSRO, and QRO, indicating a potential for \textbf{DNS exfiltration}. Indeed, our post-hoc analysis showed that on 30-Jun it sent out $709K$ DNS queries with pattern {\myverb{SARICA[10digits].com}} towards an IP address in Turkey, and on the next day, another $964K$ DNS queries to the same server with pattern {\myverb{akbank[9digits].com.tr}} -- the random 9 or 10 digits very likely encode exfiltrated data as highlighted in \cite{JAhmedTNSM2020}.

\begin{table}[t!]
	\caption{DNS anomalies considered in this paper and their indicative alerts and required post-hoc analysis.}
	\centering
	\label{tab:InferDNSAnomalies}
\begin{adjustbox}{max width=0.48\textwidth}
	\renewcommand{\arraystretch}{1.2}	
	\begin{tabular}{|l|c|c|}
		\hline
		\rowcolor[rgb]{ .906,  .902,  .902}	\textbf{ DNS Anomaly Type}&	\textbf{Indicative Alerts}   & \textbf{Post Analysis}  \\ \hline\hline
		\textbf{A1}: Misconfiguration & $\uparrow$ NELF \& LEF & None\\ \hline
		\textbf{A2}: Query DDoS & $\uparrow$ QSRI \& QRI & Flow profile\\ \hline
		\textbf{A3}:Response DDoS & $\downarrow$ QSRO \& $\uparrow$ RRI & Flow profile\\ \hline
		\textbf{A4}: Attack reflector & $\uparrow$  QRI \& RRO & Flow profile\\ \hline
		\textbf{A5}: Generating scan & $\uparrow$ LEF \& $\downarrow$ QSRO & Flow profile\\ \hline
		\textbf{A6}: Data exfiltration & $\uparrow$ LEF \& QRO  \& $\downarrow$ QSRO&  Query content   \\ \hline\hline
		\textbf{A4'}: Reflector (after fix)& $\uparrow$ QRI \& RRO &Flow profile\\ \hline
	\end{tabular}
\end{adjustbox}
\vspace{-3mm}
\end{table}

\textbf{Alerts across the two organizations:}
Certain DNS assets -- $35\%$ in the University and $13\%$ in the research institute -- were consistently flagged by alerts in each epoch. %(shown as the right tails of both distributions in {\color{red}Fig.~\ref{fig:hostHealth}}). 
These turn out to be largely Authoritative DNS servers that are publicly facing, and hence exposed to inbound DNS attacks (interesting, most of these were managed by sub-departments or third-parties, rather than central IT in the organization). Recursive resolvers in both organizations raised relatively fewer alerts, typically in {\myverb{QRO}} and {\myverb{RRI}} during some epochs. 

\begin{figure}[t!]
	%	\hspace{-5mm}
	%		\vspace{-10mm}
	\begin{center}
		%			\vspace{-2mm}
		\mbox{
			\hspace{-7mm}
			%		\vspace{-3mm}
			\subfigure[Univesity campus.]{
				\hspace{-2mm}
				{\includegraphics[width=0.255\textwidth]{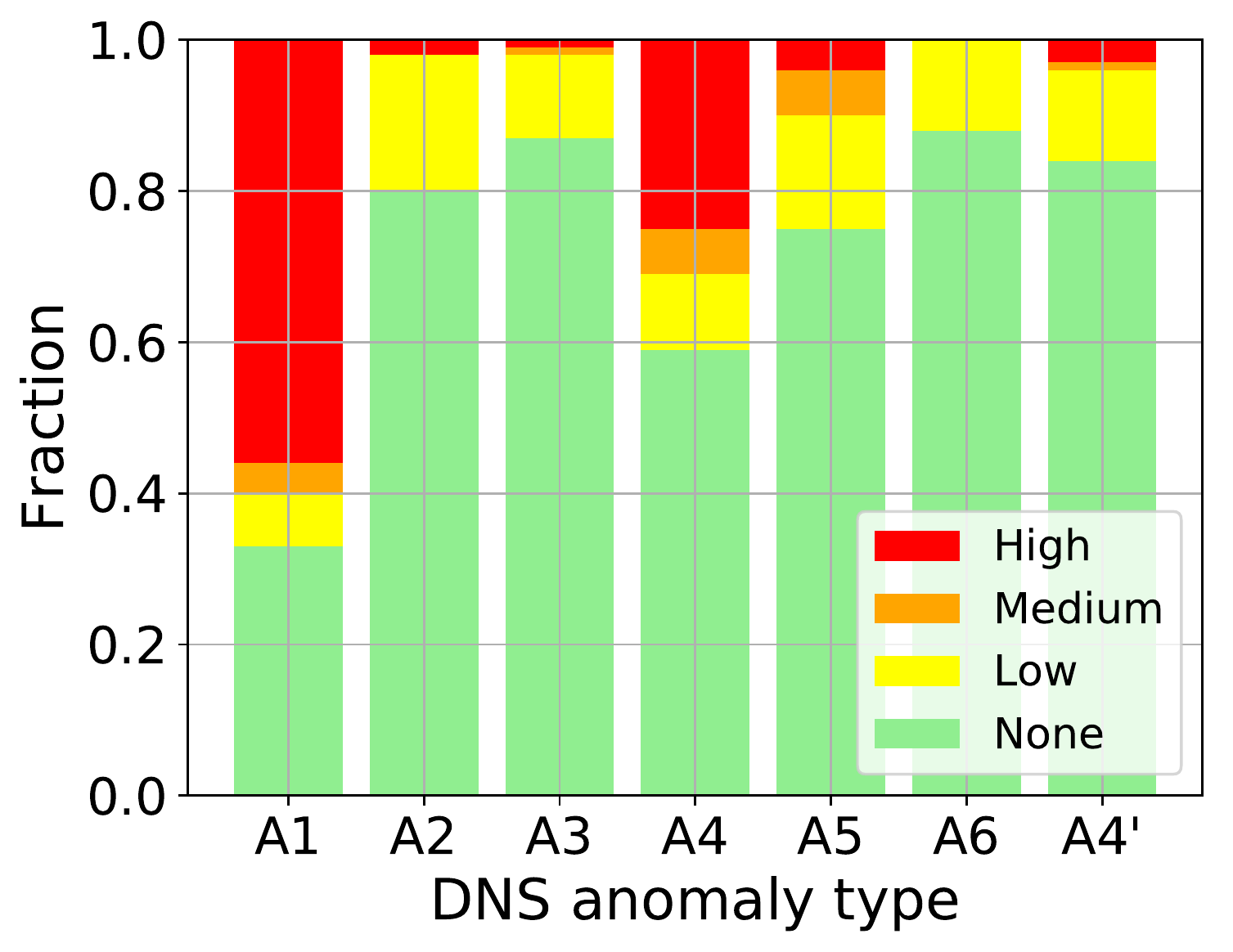}}
				\label{fig:AnomalyFractionUniv}
				%		\vspace{-10mm}
			}
			%		}
			%		\hspace{-15mm}
			%		\mbox{
			%	\hspace{-5mm}
			%		\vspace{-8mm}
			\subfigure[Research institute.]{
				\hspace{-6mm}
				{\includegraphics[width=0.255\textwidth]{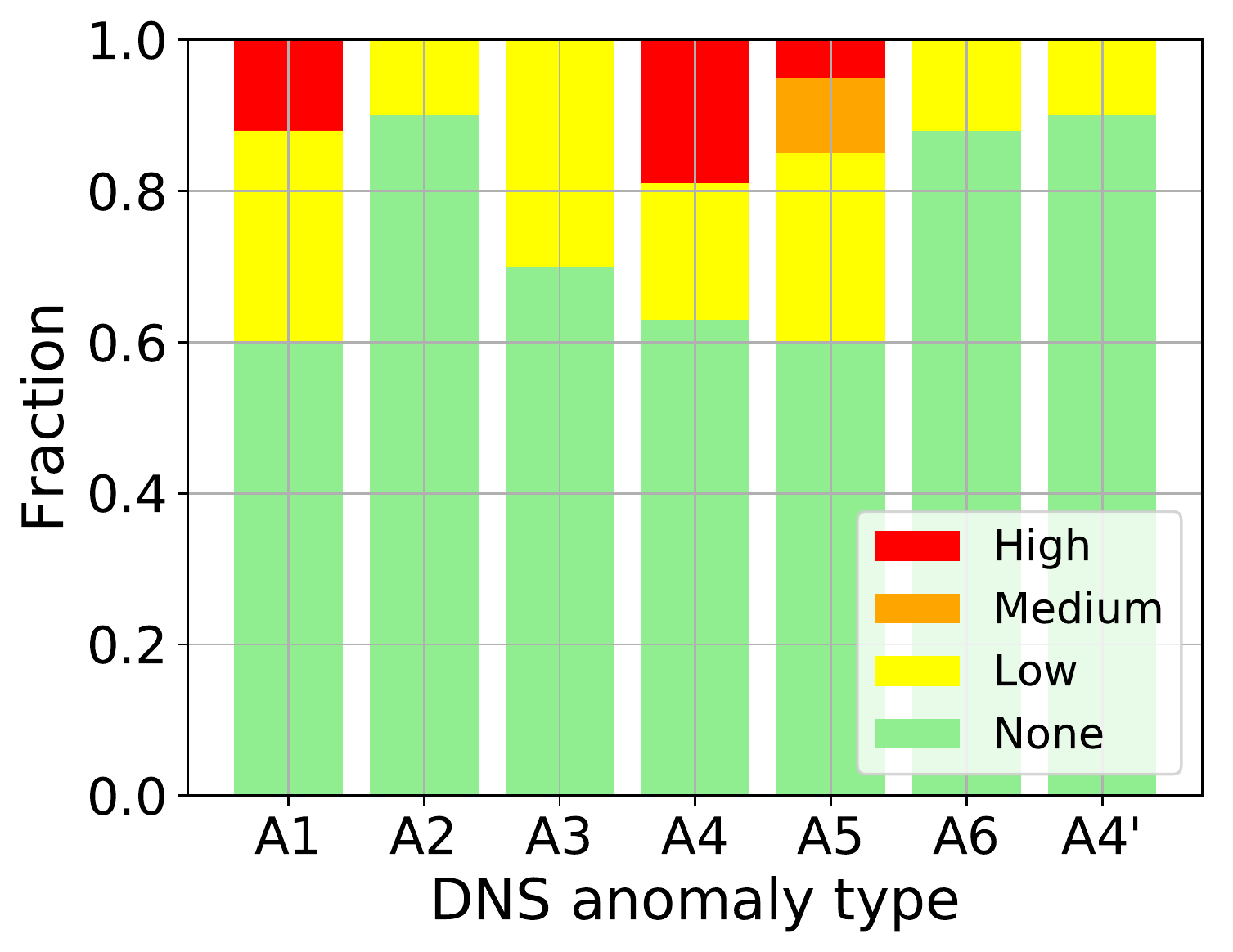}}
				\label{fig:AnomalyFractionRsrch}
			}
		}
		\vspace{-3mm}
		\caption{Severity of DNS anomalies of each enterprise asset in both organizations.}
		\vspace{-5mm}
		\label{fig:AnomalyFraction}
	\end{center}
		\vspace{-2mm}
\end{figure}

\textbf{Inferring Anomalies from Alerts:}
Tracking the health metrics (aka ``symptoms'') allows us to make inferences about the underlying anomalies (aka ``diseases''). We built a simple inference engine using the Codebook Correlation technique used extensively in Network Management for event correlation \cite{SKligerINM1995}. A causality graph was built as per Fig.~\ref{fig:MappingAnomalyTypeTrafficHealth}, a codebook correlation model was derived, and then ``alerts'' from the 32-day dataset were looked up in the codebook to determine the underlying ``anomaly''. The outcomes, in terms of the health of the DNS assets across the two organizations, are shown in Fig.~\ref{fig:AnomalyFraction}. 

%1. the major problem -- misconfig.
Our first observation is that misconfiguration is a significant problem across both organizations -- $56\%$ and $33\%$ of DNS assets in the University and research institute, respectively, serve DNS queries not relevant to the enterprise. This is a serious concern -- Authoritative DNS servers are resolving non-enterprise queries and thereby being exposed to random queries, which can lead to denial-of-service; while recursive resolvers are resolving queries for non-enterprise hosts, thus being made available to attackers as reflectors for DDoS attacks on spoofed victims. Indeed, our analysis shows that if these DNS configurations were to be rectified, the number of DNS assets being used as reflectors falls from $25\%$ to $3\%$ in the University, and from $20\%$ to $0\%$ in the Research Institute (shown as the rightmost bars of Fig.~\ref{fig:AnomalyFractionUniv} and \ref{fig:AnomalyFractionRsrch}).

%2. Scans
The second most significant concern is that there is evidence of scans emanating from both organizations, as indicated by epochs of high lookup failures (LEF) and low success of responses (QSRO). These indicate that there is malware lurking within the organizations that is using DNS to perform scans on other Internet hosts. Identifying malware-infected hosts would require access to traffic within the organization, which is beyond the scope of this paper. 

Finally, we note that there are epochs with evidence of DNS data exfiltration from the University network. Again, knowing the hosts complicit in this requires analysis of traffic within the organization (our traffic feed at the border does not tell us which internal host made the DNS request to the organizational cache/proxy), which is beyond the scope of this paper. Similarly, a few assets in the Research Institute are occasionally launching DDoS attacks on external victims.

While we do not intend to diagnose every DNS problem, it is continuously assessing the health of each DNS asset in the organization, and flagging potential issues that can be investigated further by the network operator, providing them actionable intelligence to rectify misconfigurations, amend firewall policy rules, rate-limit query rates, etc., to better protect their assets.

\vspace{-3mm}
\section{Conclusion}\label{sec:concl}
%\vspace{-3mm}
Enterprise networks are often vulnerable to DNS-based cyber attacks due to insufficient monitoring of DNS traffic. In this paper, we have developed methods to classify enterprise assets and continuously track their cyber-health by passively analyzing DNS traffic crossing the network border of an organization.
We performed a comprehensive analysis of DNS packets from two large organizations to identify asset profiles by network, functional, and service characteristics. We highlighted the behavior of enterprise hosts, either benign and anomalous.
We then trained unsupervised machine learning models by DNS traffic attributes that classify the DNS assets, including authoritative name server, recursive resolver, mixed DNS server, and end-hosts behind or not behind the NAT.
Lastly, we developed metrics to track the cyber health of enterprise DNS assets continuously. We identified several instances of improper configurations, data exfiltration, DDoS, and reflection attacks. 
Results of our real-time application have been verified with IT departments of the two organizations while revealing unknown knowledge that helps them enhance their security management without incurring risks and excessive labor costs. 

%\vspace{-3mm}

\bibliographystyle{IEEEtranS}
\bibliography{References}
\balance

\end{document}